\documentstyle[prb,aps,psfig,multicol
]{revtex}
\begin{document}
\draft
\title{Thermodynamics of a 4-site Hubbard model by 
analytical diagonalization}
\author{R. Schumann}
\address{Technical University Dresden, Institute for Theoretical 
Physics,\\ D-01062 Dresden, Germany}
\date{\today}
\maketitle
\begin{abstract}
By use of the conservation laws a four-site Hubbard model coupled to a particle
bath within an external magnetic field in z-direction was diagonalized. The
analytical dependence of both the eigenvalues and the eigenstates  
on the interaction strength, the chemical potential and magnetic
field was calculated. 
It is demonstrated that the low temperature behaviour is determined by a delicate interplay
between many-particle states differing in electron number and spin if the
electron density is away from half-filling. 
The grand partition sum is calculated  and the specific heat, 
the suszeptibility as well as various correlation functions and spectral
functions are given in dependence of the interaction strength, the electron
occupation and the applied magnetic field. 
Furthermore, for both the grand canonical and the canonical ensemble
the so called crossing points of the suszeptibility are calculated. They
confirm the universal value predicted by Vollhardt \cite{Vollhardt97}.
\end{abstract}


\begin{multicols}{2}
%
\newcommand{\beq}{\begin{eqnarray} \displaystyle}
\newcommand{\eeq}{\end{eqnarray}}
\newcommand{\summe}[1]{\sum\limits_{#1}}
\newcommand{\sumijs}{ \sum\limits_{i\neq j\sigma}}
\newcommand{\product}[1]{\prod\limits_{#1}}
%
\newcommand{\cplus}[1]{{\bf c}^+_{#1}}
\newcommand{\cminus}[1]{{\bf c}_{#1}}
\newcommand{\cpup}[1]{{\bf c}^+_{#1 \uparrow}}
\renewcommand{\cup}[1]{{\bf c}_{#1 \uparrow}}
\newcommand{\cpdown}[1]{{\bf c}^+_{#1 \downarrow}}
\newcommand{\cdown}[1]{{\bf c}_{#1 \downarrow}}
\newcommand{\cpsigma}[1]{{\bf c}^+_{#1 \sigma}}
\newcommand{\csigma}[1]{{\bf c}_{#1 \sigma}}
\newcommand{\cpis}{{\bf c}^+_{i \sigma}}
\newcommand{\cis}{{\bf c}_{i \sigma}}
\newcommand{\cpjs}{{\bf c}_{j \sigma}}
\newcommand{\cjs}{{\bf c}_{j \sigma}}
\newcommand{\apnull}{{\bf a}^+_{0}}
\newcommand{\anull}{{\bf a}_{0}}
\newcommand{\apup}{{\bf a}^+_{\uparrow}}
\newcommand{\aup}{{\bf a}_{\uparrow}}
\newcommand{\apdown}{{\bf a}^+_{\downarrow}}
\newcommand{\adown}{{\bf a}_{\downarrow}}
\newcommand{\apzwei}{{\bf a}^+_{2}}
\newcommand{\azwei}{{\bf a}_{2}}
\newcommand{\apsigma}[1]{{\bf a}^+_{#1 \sigma}}
\newcommand{\asigma}[1]{{\bf a}_{#1 \sigma}}
\newcommand{\aplus}[1]{{\bf a}^+_{#1 }}
\newcommand{\aminus}[1]{{\bf a}_{#1 }}
%
%
\newcommand{\nup}[1]{{\bf n}_{#1 \uparrow}}
\newcommand{\ndown}[1]{{\bf n}_{#1 \downarrow}}
\newcommand{\nsigma}[1]{{\bf n}_{#1 \sigma}}
\newcommand{\noperator}[1]{{\bf n}_{#1 }}
\newcommand{\dyade}[2]{| #1 \rangle \langle #2|}
\newcommand{\hamilton}{{\bf H}}
\newcommand{\gesamtn}{\mbox{\boldmath $N$}}
\newcommand{\gesamtr}{\mbox{\boldmath $S$}}
\newcommand{\gesamts}{\mbox{\boldmath $R$}}
\newcommand{\liouville}{\mbox{\boldmath $L$}}
\newcommand{\interaction}{\mbox{\boldmath $V$}}
\newcommand{\koperator}{\mbox{\boldmath $K$}}
\newcommand{\operator}[1]{{\bf #1}}
\newcommand{\opP}{\mbox{\boldmath$ P $}}
\newcommand{\opQ}{\mbox{\boldmath $ Q $}}
\newcommand{\opA}{\mbox{\boldmath $ A $}}
\newcommand{\opB}{\mbox{\boldmath $ B $}}
\newcommand{\opC}{\mbox{\boldmath $ C $}}
\newcommand{\opD}{\mbox{\boldmath $ D $}}
\newcommand{\opL}{\mbox{\boldmath $ L $}}
\newcommand{\opH}{\mbox{\boldmath $ H $}}
\newcommand{\ri}[1]{{\bf R}_{i #1}} 
\newcommand{\si}[1]{{\bf S}_{i #1}}
\newcommand{\eins}{\mbox{\boldmath$I$ }}
\newcommand{\einsa}{\mbox{\boldmath$I\!_2$ }}
\newcommand{\laplace}{\mbox{\boldmath$\Delta$ }}
\newcommand{\opbeta}{\mbox{\boldmath $ \beta $ }}
\newcommand{\Sigmaz}{\mbox{\boldmath $ \Sigma_z$}}
\newcommand{\sigmax}{\mbox{\boldmath $ \sigma_x$}}
\newcommand{\sigmay}{\mbox{\boldmath $ \sigma_y$}}
\newcommand{\sigmaz}{\mbox{\boldmath $ \sigma_z$}}
\newcommand{\opsigma}{\vec{\mbox{\boldmath $ \sigma $}}}
\newcommand{\opalpha}{\vec{\mbox{\boldmath $ \alpha $}}}
\newcommand{\tket}[1]{$ | #1 \rangle$}
\newcommand{\tbra}[1]{$\langle #1 |$}
\newcommand{\mket}[1]{| #1 \rangle}
\newcommand{\mbra}[1]{\langle #1|}
\newcommand{\ket}[1]{| #1 \rangle}
\newcommand{\bra}[1]{\langle #1|}
\newcommand{\scalar}[2]{\langle #1 \, | \, #2 \rangle}
\newcommand{\expect}[1]{\langle #1 \rangle}
\newcommand{\texpect}[1]{$\langle #1 \rangle$}
\newcommand{\tmean}[1]{$\langle #1 \rangle$}
\newcommand{\ketvacuum}{\mket{\mbox{vac}}}
\newcommand{\bravacuum}{\mbra{\mbox{vac}}}
\newcommand{\tpket}[1]{$ \left| #1 \right )$}
\newcommand{\tpbra}[1]{$ \left ( #1 \right|$}
\newcommand{\pket}[1]{\left | #1 \right )}
\newcommand{\pbra}[1]{\left ( #1 \right |}
\newcommand{\pscalar}[2]{\left ( #1 
\,\left | #2 \right . \right )}
\newcommand{\ikr}{{\rm e}^{i\vec{k}\vec{R}_i}}
\newcommand{\mikr}{{\rm e}^{-i\vec{k}\vec{R}_i}}
\newcommand{\ikvonr}[1]{{\rm e}^{i\vec{k}\vec{R}_{#1}}}
\newcommand{\mikvonr}[1]{{\rm e}^{-i\vec{k}\vec{R}_{#1}}}
\newcommand{\ryd}{\mbox{ryd}}

\newcommand{\sumviervier}{\summe{k \alpha} \summe{k' \alpha'} 
                   \summe{q \beta} \summe{q' \beta'}}
\newcommand{\sumvier}{\summe{k \alpha} \summe{k' } 
                   \summe{q \beta}  \summe{q' }}                     
\newcommand{\sumsechs}{\summe{k \sigma} \summe{k' } 
                   \summe{q}  \summe{q' }
                   \summe{p}  \summe{p' }}                     
\newcommand{\sumkkqq}{\summe{k } \summe{k' } 
                   \summe{q }  \summe{q' }}                     
\newcommand{\sumijlm}{\summe{i \sigma} \summe{j } 
                   \summe{l}  \summe{m }}                     
\newcommand{\sumij}{\summe{i \sigma} \summe{j }} 
\newcommand{\ddelta}[1]{\delta_{#1,0}}
\newcommand{\expo}[1]{e^{\displaystyle #1}}
\newcommand{\difk}{\left(\frac{\vec{p}}{|\vec{p}|} \frac{\partial}{\partial \vec{k}}\right)}
\newcommand{\difq}{\left(\frac{\vec{p}}{|\vec{p}|} \frac{\partial}{\partial \vec{q}}\right)}
\newcommand{\diffk}{\mbox{\bf D}}
\newcommand{\difkx}{p_x \frac{\partial}{\partial k_x}}
\newcommand{\difky}{p_y \frac{\partial}{\partial k_y}}
\newcommand{\diffkx}{\mbox{\bf D}_x}
\newcommand{\diffky}{\mbox{\bf D}_y}
\newcommand{\rtot}{{\bf R}}
\newcommand{\stot}{{\bf S}}
\newcommand{\imag}{i}
\newcommand{\Aindex}[1]{A_{#1}}
\newcommand{\Bindex}[1]{B_{#1}}
\section{Introduction}
The Hubbard model \cite{Hubbard63,Gutzwiller63} with the restraint
of electron hopping to nearest neighbours only is certainly one of the 
simplest many particle models. It was utilized for the description of 
correlated electrons in the intermediate and strong coupling regimes. 
First it was used in chemistry to
describe unsaturated hydrocarbons \cite{Pariser53,Pople53}. In solid state
physics it was introduced to
describe the behaviour of narrow-band transition metal compounds, with the
main emphasis given to the explanation of the metal-insulator transition
\cite{Hubbard63,Gebhard97,Imada98}
and the itinerant ferromagnetism \cite{Moriya}.
Later on it played a central role in the heavy fermion context, and especially
the discovery of high T$_c$ 
superconductivity in the cuprates renewed the interest, since its involved 
spin dynamics is
widely considered as a good candidate to understand at least qualitatively
the pairing mechanism in the copper oxide planes, being common to all high T$_c$
superconductors. It is clear that after about four decades of research
there exist a vast literature and meanwhile also some good
reviews dealing with different aspects of the model 
\cite{FuldeBuch,Rasetti91,Montorsi92}, 
where good reasons to
deal with that model are given, so I refer to them. In contrast to the
large amount of work spent to that model, there is astonishing few 
exact knowledge. The initial hope that by reducing the description
of the correlated electrons to a one-band model with hopping restricted
to next nearest neighbours and correlation restricted to the on-site
term a model will result, which can be solved analytically, has not yet been
fulfilled in general.
Nevertheless, Lieb and Wu \cite{Lieb68} calculated the ground-state-energy of
the one dimensional model with a filling of one electron per lattice site by help 
of the Bethe ansatz. 
Starting from this paper there were later on attempts to get the complete
spectrum and thus the thermodynamics for the half filled model and during the 
last decade also for
arbitrary electron filling, at least approximately. 
The state of art for the one dimensional
model was recently summarized by Deguchi et. al. \cite{Deguchi99}.
All theories starting from the Bethe ansatz result in a set of coupled
nonlinear equations, the Lieb-Wu equations, which have to be solved
numerically for every parameter set. Thus all studies of physical
quantities like susceptibilities or excitation spectra resting upon the
solution of the Lieb-Wu equations have to be done numerically, which often
limits the application. The transfer matrix method introduced by Shastry 
\cite{Shastry86} supplies an alternate solution, which is in principle exact.
The drawback for calculating physical quantities is that the
lowest eigenvalue of the transfer matrix has to be calculated, what
requires again a considerable numerical effort. For the cases where results
exist,
they were shown to be equivalent 
to the Bethe ansatz method \cite{Deguchi99}.
Another way to achieve exact information is the numerical treatment of finite
systems. The techniques by far most intensively used for this task are
the Lanczos algorithm \cite{Dagotto84} and Monte Carlo methods \cite{Binder92}.
Lanczos calculations. Albeit giving essentially exact results for
ground-state and low lying eigenvectors and single particle excitation
spectra both methods suffer from the relative small cluster size treatable with
the contemporary computer facilities, since the basis set of vectors
grows exponentially with the system size. The main drawback in the MC
calculations for fermions are the relatively large uncertainties due
to the so called sign problem, although there exist a enormous number of
derivate methods developed to overcome this problem. 
Direct exact diagonalization, e.g. with the householder method, is
restricted to much smaller problems and exact is then always meant numerically.
Along this way
it is in principle impossible to get closed analytical expressions in
dependence of the model parameters.
In most cases a numerically exact calculation may be enough, but there are 
problems which need the explicit analytical
dependence of the eigenvalues and the coefficients of the eigenvectors with 
respect to a suitable basis set. The problem of determining the crossing 
temperature of the high temperature specific heat, a problem we address in
the third section, is such a case, since it invokes the third derivative of the
thermodynamical potential. Another problem, where the analytical knowledge
of the spectrum is inevitable, is the term-by-term comparison
of series expansions with exact solutions, and, last but not least, the
study of the influence of a multitude of external parameters,
(e.g. in the model
under consideration the ratio t/U between the hopping energy and the on-site
correlation energy, the electron density n, the temperature, and the external
applied fields) is considerably
simplified, if the eigensystem of an Hamiltonian is known analytically,
especially, if the calculation aims at the grand canonical partition sum, where
the chemical potential has to be determined from the particle number.
To get the explicit dependencies on the model parameters it
is necessary to diagonalize the Hamiltonian in an analytical manner, 
what restricts the calculation to very small systems of course. 
We have chosen a four-point square for this task for the following reasons: 
(i) It is the earliest 
\cite{Heinig72a,Heinig72b,Shiba72,Cabib73,Roessler81,Newman84}
and most often studied non-trivial object in numerical diagonalization.
(ii) There were some attempts in the nineties which remained incomplete, since
they discussed the half filled case only, where the eigenvalues were given. 
The authors speculate about hidden symmetries, since they were not able to
resolve all degeneracies \cite{Villet90,Noce96}.
(iii) It is a special case of the N-site rings,
which are subject of the above cited Bethe ansatz and quantum transfer matrix
methods, and might be useful as a proving tool. 
(iv) In the following sense the four point Hubbard square may serve as a
crude model to the square lattice: If you neglect every second
hopping in x- and y-direction respectively you will get a system of N/4 
independent 4-site-clusters being the topic of this paper, see
Fig. \ref{clustergas}.
The model is thus
a little bit better than the so-called atomic limit, were every hopping is
neglected. The main fault of this cluster-gas model is that there is no direct
electron transfer between the clusters. This is in some sense  corrected 
by applying the grand canonical distribution, since it reflects the situation
where the subsystems are in contact with a particle reservoir allowing 
for particle number fluctuations, thus establishing an indirect particle
transfer via the bath. 
Very recently the spectral function of the square-lattice Hubbard model
was calculated by a cluster-perturbation method,  
which used the square-clusters as unperturbed system and the
inter-square-hopping as perturbation \cite{Senechal00}, albeit they 
diagonalized the cluster numerically. 
(iv) The rigorous solution of any strong interacting many particle system 
is a value independent of its applicability to realistic systems, 
at least for pedagogical reasons. 
(v) The limited resources of the available computers, we used for symbolic
formula handling. 

The paper is organized as follows. 
In the next two sections the diagonalization procedure is explained and the
many-particle spectrum is discussed. In the fourth section we give the complete
thermodynamics, based on the grand canonical potential. The main focus lies on
the dependence of the physical quantities on the electron density, whereas the comparison of
our results to the literature for the half-filled case serves as independent
proof. In the closing section we discuss, what we could learn from our
analytical solution.  
\section{The analytical diagonalization procedure}
We are interested in
the eigenvalue sequence for the operator 
\beq
\!\!\!\!
\hamilton &=&  t \sumijs \cplus{i \sigma}\cminus{j \sigma} + 
\summe{i \sigma} \left(
\frac{U}{2}\nsigma{i}\noperator{i-\sigma}-(\mu+\sigma h) \nsigma{i} \right )
\label{hamilton}
\eeq
Here $\cplus{i \sigma}$ and $\cminus{i \sigma}$ are the creation and 
destruction operators
in Wannier representation. 
The chemical potential $\mu$  and the magnetic 
field h in z-direction, which we choose to be normal to the square, are introduced
to take into account the effects of doping and applying external magnetic fields. 
The related spectrum we will call grand-canonical, 
due to its relation to the
weights within the grand canonical distribution. The lowest level of this
operator determines the state with the highest weight, becoming one for
$T \rightarrow 0$, consequently, we will call it the grand canonical
groundstate, which of course may be different for different values of
$\mu$ and h.  
The many particle Hilbert space is the direct product of the
four single-site Hilbert spaces, each containing four states, i.e. the
empty state, the two single occupied states with spin up and spin down, 
and the double occupied state. Thus we have 256 states. For example, $\ket{u,0,2,d}$
indicates the state which has a spin-up electron at the
first point, no electron at the second position, two electrons at the third,
and one spin-down electron at the last position.
Of course it is impossible to diagonalize the resulting
256x256 matrix in an analytic manner straight forwardly, instead one has 
to use all the known symmetries or the related conservation laws to reduce 
the matrix to block diagonal
form with all blocks being at least 4x4. 

The Hubbard model is very good suited
for the task of analytical diagonalization by help of the conservations laws, 
since it contains several conserved quantities which are easy to handle.
First of all, there are the two commuting SU[2] algebras of the total spin and 
the total pseudospin \cite{Heilmann71} providing us with four operators
commuting with each other and with the hamiltonian, i.e.
\beq
[ \stot^2, \hamilton] = [ \stot_z, \hamilton] = [ \rtot^2,\hamilton]= [\rtot_z, \hamilton 
] = 0
\eeq
with
\beq
\stot^2&=& \frac{1}{2}\left(\stot^+\stot^- + \stot^- \stot^+ \right) + \stot_z^2 \\
\stot^+&=& \sum_i \cpup{i} \cdown{i} \quad,\quad\\
\stot^-&=& \sum_i \cpdown{i} \cup{i} \quad,\quad\\
\stot_z&=& \sum_i \frac{1}{2}\left( \nup{i} - \ndown{i} \right) \\
\rtot^2&=& \frac{1}{2}\left(\rtot^+\rtot^- + \rtot^- \rtot^+ \right) + \rtot_z^2 \\
\rtot^+&=& \sum_i (-1)^i \cpup{i} \cpdown{i} \quad,\quad\\
\rtot^-&=& \sum_i (-1)^i \cdown{i} \cup{i} \quad,\quad\\
\rtot_z&=& \sum_i\frac{1}{2}\left( \nup{i}  + \ndown{i} -1 \right )
\eeq
We mention here that the total pseudospin is conserved only if the hopping
matrix elements fulfill the condition 
\beq
t_{ij}\left( (-1)^i+(-1)^j \right) = 0
\eeq
what is the case if one restricts e.g. the hopping to next neighbours as we
supposed in eq. (\ref{hamilton}).
The expectation value of $\rtot_z$ is nothing but the difference of the model
from the half filled case, or in the context of high-$T_c$ superconductivity
it is the doping of the model. It can be controled by adjusting the chemical
potential $\mu$ of course.
Dealing with a special cluster one can use the symmetries of it. In our case
the actual symmetry is C$_{4v}$. Indeed there are
two works on the four-site square, however restricted to the half filled case,
which tried to use these symmetries. Within
the first \cite{Villet90} the authors start from a parametric expression of
the Hubbard interaction, which they took obviously from Ref.~\onlinecite{Schumann89}.
But instead of using both spin and pseudospin symmetry they used the spin
symmetry first and then the point symmetry of the four site cluster. This way
the biggest block in the hamiltonian remains 7x7, what in general prevents 
an analytical solution. Nevertheless they found the roots. More elegantly the
same task was done in Ref.~\onlinecite{Noce96}, where both spin and
pseudospin were used first and the translation invariance afterwards. 
This is enough to
bring the hamiltonian matrix being 70x70 in the case of the half filled model
into the wanted block form with none of the remaining 
blocks being bigger than 4. In both papers the authors mentioned a degeneration
of the energy levels, which they ascribed to some hidden symmetries, a point
which will be clarified within the present work. \\
In this paper we use the same method for arbitrary 
filling. Due to the commutation of $\rtot_z$ and $\stot_z$ with the hamiltonian,
we started with the basis of the Hilbert subspace with a given value of $\rtot_z$
or, equivalently, with a given number  of electrons, and a given value of
$\stot_z$. Afterwards we calculated the matrix of $\rtot^2$ and diagonalized it
resulting in a common eigensystem of $\rtot_z , \stot_z , \rtot^2 $. The next
step was the diagonalization of $\stot^2$ in every subspace for given
eigenvalues $\rtot_z , \stot_z , \rtot^2 $ thus getting  common eigenvectors to
$\rtot_z , \stot_z , \rtot^2, \stot^2 $. The last step, if necessary, was the
calculation of the matrix of the translation operator, which fulfills $\operator{U}^4
\ket{\mu}=\ket{\mu}$
for a four-site ring  and therefore has the eigenvalues
1, i, -1, and -i. We took the operator $\operator{U}$ 
in the form given in Ref.~\onlinecite{Goehmann}.
\beq
\operator{U}&=&\operator{U}_{\uparrow}\operator{U}_{\downarrow}\\
\operator{U}_{\sigma}&=& \operator{U}_{12\sigma}\operator{U}_{23\sigma}
\operator{U}_{34\sigma}\\
\operator{U}_{ij\sigma}&=& 1-(\cpsigma{i}-\cpsigma{j})(\csigma{i}-\csigma{j})
\eeq
The operators $\operator{U}$, $\operator{U^2}$, $\operator{U}^3$ are
operator representations of the group elements
C$_{4z}$, C$_{2z}$, and C$_{4z}^{-1}$, which form together with the 1 standing
for E an Abelian subgroup. The relation to the conserved momentum is obvious.
The operator $\operator{U}_{ij\sigma}$ permutes two
electrons and is therefore well-suited to construct the missing group
elements. We find for
\beq
\operator{P}(IC_{2x}) &:=& \operator{U}_{14\uparrow}\operator{U}_{23\uparrow}
\operator{U}_{12\downarrow}\operator{U}_{34\downarrow}  \\     
\operator{P}(IC_{2y}) &:=& \operator{U}_{14\uparrow}\operator{U}_{24\uparrow}
\operator{U}_{12\downarrow}\operator{U}_{34\downarrow}  \\     
\operator{P}(IC_{2a}) &:=& \operator{U}_{13\uparrow}\operator{U}_{13\downarrow}  \\     
\operator{P}(IC_{2b}) &:=& \operator{U}_{24\uparrow}\operator{U}_{24\downarrow}  
\eeq
Since these operators neither commute with each other nor with
the $\operator{U}$-operators we have to construct new operators commuting
with all the other operators.  The result is
\beq
\operator{P}^{++}&=&\operator{P}(IC_{2x})-\operator{P}(IC_{2y})\nonumber\\
            &&     +\operator{P}(IC_{2a})-\operator{P}(IC_{2b})\\    
\operator{P}^{+-}&=&\operator{P}(IC_{2x})-\operator{P}(IC_{2y})\nonumber\\
            &&      -\operator{P}(IC_{2a})+\operator{P}(IC_{2b})\\    
\operator{P}^{-+}&=&-\operator{P}(IC_{2x})+\operator{P}(IC_{2y})\nonumber\\
            &&      +\operator{P}(IC_{2a})-\operator{P}(IC_{2b})\\    
\operator{P}^{--}&=&-\operator{P}(IC_{2x})+\operator{P}(IC_{2y})\nonumber\\
            &&      -\operator{P}(IC_{2a})+\operator{P}(IC_{2b})
\eeq
Grosse \cite{Grosse89} constructed further conserved quantities for the one-dimensional system 
with periodic boundary conditions, which are in general highly nontrivial
due to their dependence on the interaction strength. The operator given in 
the first theorem of Ref.~\onlinecite{Grosse89} reads
\beq
\operator{K}^1&=&\operator{K}^1_0 + \frac{U}{t} \operator{K}^1_1\\
\operator{K}^1_0&=&\sum_{i,\sigma} \left( \cpsigma{i+2} \csigma{i}
-\cpsigma{i} \csigma{i+2} \right) \\
\operator{K}^1_1&=&\sum_{i,\sigma}
\left(\cpsigma{i+1} \csigma{i} - \cpsigma{i} \csigma{i+1} \right) 
\times \nonumber\\
&&\qquad \times \left(\cplus{i,-\sigma} \cminus{i,-\sigma} 
+ \cplus{i+1, -\sigma} \cminus{i+1, -\sigma}  \right)  
\eeq
For our four-site ring $\operator{K}^1_0$ is zero. In Ref.~\onlinecite{Grosse89}
there another operator $\operator{K}^2$ is given, which does not give new
results due to our short periodicity length. As mentioned above, for
our small system it is not necessary to use $\operator{K}^1$, but, 
since it commutes
with all the other operators, it must have the same eigensystem like
the hamiltonian. 
Since it  is an antihermitian operator the eigenvalues have to be zero or 
pure imaginary. Due to its rather complicated character we used it among others 
as a powerful tool for checking the eigensystem. 
All the calculations were done
by help of algebraic programming. The resulting eigenvectors and the related
eigenvalues were proved by help of the eigenvalue equation. Also the matrices
for the special case $N_e=4$ are compared to Ref.~\onlinecite{Noce96}, where they are given. The 
matrices $\hamilton_{0,0,1}$ and $\hamilton_{1,0,1}$ \cite{Noce96} 
differ in the sign for two
off-diagonal matrix elements, but the eigenvalues are the same and the
matrices $\hamilton_{0,0,-1}$ and $\hamilton_{0,1,1}$ are interchanged. Since the
resulting spectrum for the half filled case is not altered by these
differences, the conclusions drawn in Ref.~\onlinecite{Noce96} remain true.
The complete set of the eigenvalues and eigenvectors \tket{m_r,
r(r+1), m_s, s(s+1), u, k_{11}, p_{++}, p_{+-}}, with $m_r, m_s, s, r, u,
k_{11},p_{++},p_{+-}$ indicating the eigenvalues of the operator-set 
$\rtot_z , \rtot, \stot_z, \stot^2 , \operator{U}, \operator{K}_{11},
\operator{P}^{++}, \operator{P}^{+-} $, in its analytical dependence 
on the model parameters are given in the appendix. 
Inspecting the spectrum carefully we found all the states classified by
the eigenvalues of the conserved quantities in a unique manner. Thus we
clarified the reason of the mysterious degeneracies mentioned in 
Ref.~\onlinecite{Villet90}. Also in Ref.~\onlinecite{Noce96} hidden degeneracies remained,
although not mentioned by the authors. For instance the two degenerate states 
No. 95 and 96 have identical eigenvalues of the conserved quantities 
$\rtot_z , \rtot, \stot_z, \stot^2 , \operator{U}$, which was the set exploited
in Ref.~\onlinecite{Noce96}. This degeneracy is due to the conserved 
Grosse operator $\operator{K}_{11}$. 
Heilmann and Lieb \cite{Heilmann71} showed for the benzene that
there are a multitude of level crossings. Since they have constructed
all correlation-independent symmetries they ascribe this fact to the
existence of unknown U-dependent symmetries. We believe, that the Grosse
operators \cite{Grosse89} are candidates for the missing symmetry, since they were
left out in Ref.~\onlinecite{Heilmann71}. For instance, it is obvious that for a six-site
ring the operator $\operator{K}^1$ is U-dependent. Nevertheless, 
in our special case it is not, 
since the operator $\operator{K}_0^1$ vanishes. Thus within the four-site
model $\operator{K}_1^1$  represents a further U-independent symmetry.
By this way we remark that there is no contradiction to the non-crossing
rule, since all the eigenstates differ in the symmetry quantum numbers
of mutual commuting operators, a question raised in Ref.~\onlinecite{Heilmann71}. 
\section{The grand canonical eigensystem}
\label{diagonal}
\subsection{The low lying states} 
In the following discussion of the spectrum we always mean the grand 
canonical levels, which are numbered for referencing according to the appendix,
exceptions are explicitly indicated.
The well-known particle-hole symmetry allows to fix the electron number to 
four if we choose $\mu$=U/2. 
Looking at the lower part of the spectrum of the half filled model, 
as plotted in Fig. \ref{spectrumU5N4h01}, 
we find the state No. 111 being the groundstate. 
As already mentioned in Ref.~\onlinecite{Noce96} this is 
both a spin and a pseudospin singlet (m$_r$=0 and m$_s$=0) in accordance with
the statements of  Lieb's positive-U theorem \cite{Lieb89} and the lemma of 
Shen and Qiu \cite{Shen93}.
In Fig. \ref{NviaMueU5h01} we show the results for different temperatures. 
As the first exited states we find  a spin triplet
(states No. 102, 128, 154), regarding the pseudospin it remains a singlet.
The next group consists again of a singlet-triplet type sequence, where the
singlet differs from the groundstate by its eigenvalue of the translation
operator only (it is 1 in comparison with U=-1 of the groundstate), the triplet
states have pure imaginary U eigenvalues and are therefore twice degenerated,
in agreement with a statement given in Ref.~\onlinecite{Noce96}. It is interesting, that
the next exited states deviate in the charge. Being both spin and pseudospin
doublets they are eightfold degenerated, if we do not apply neither an external 
magnetic field nor a deviation of the chemical potential from U/2. Since we
calculated Fig. \ref{spectrumU5N4h01} with a small magnetic field, 
the spin degeneracy is lifted.
We also find a pseudospin triplett (not shown in Fig. \ref{spectrumU5N4h01})
among the low lying states, i.e. the
states with No. 22 (n=2), No.131 (n=4) and No. 232 (n=6). A small deviation
of $\mu$ from U/2 will lift the degeneracy of these states, since $\mu$ couples
to $\rtot_z$ in the same manner as the magnetic field couples to $\stot_z$.
But in between the pseudospin singlet and triplet states we find a further 
pseudospin-singlet-spin-quintet-structure and pseudospin-doublet-spin-quartet
structure. It was quite surprising for us, that the low lying
spectrum of the half-filled model contains states with electron numbers 
differing from four by one and two.
For finite temperatures and deviation from half-filling
we have to determine the chemical potential from the $N(\mu)$ curves.
Due to the analytically given expression, this is in principle possible in 
any accuracy. For example, the chemical potential used in the 
Fig. \ref{spectrumU5N3h01},
where we depicted the lowest part of the spectrum for \tmean{N_e}=3, results 
in a deviation of the electron number from 3 which is less 
than $10^{-12}$ for $T=10^{-7} t$. 
If the temperature reaches zero, we have a step curve. Surprisingly
the model steps from N$_e$=2 to N$_e$=4 with no plateau at N$_e$=3.
This corresponds to
the interesting fact that there is an extreme small $\mu$-window, where
a n=3 state is groundstate. This is demonstrated by help of the inset of
Fig. \ref{GSviaMueU5h01} where we plotted the groundstate in dependence of the chemical
potential. 
Thus it is interesting to look at the spectrum with N$_e$=3 also. Since we work
with the grand-canonical potential we have to determine first the chemical
potential. 
Fig. \ref{spectrumU5N3h01} makes clear that we find again states with different 
electron numbers among the lowest levels.
The degeneration of the pseudospin-triplet is now lifted due to the large deviation
of the chemical potential from U/2, and its lowest state (No. 22) is the first
exited state above the groundstate, which is No. 46 and No. 50. 
Both states
belong to the N$_e$=3 pseudospin doublets in the half-filled model. As mentioned
above the degeneracy of the groundstate is due to the imaginary eigenvalue
of the translation operator. Among the low lying states the No. 111, 
being the groundstate of the half-filled case, remained.
If we do not apply a small magnetic field, the states No. 46 and No. 50
are degenerated with the states No. 70 and 74, and so do the states No. 22 and 
No. 111, which become the groundstates of the model for lower correlation,
although
their occupation eigenvalue is not 3.
This situation is depicted in 
Fig. \ref{spectrumU4U5N3h0low}.
Due to the interplay of the four s=1/2 states, 
belonging to N$_e$=3, and the two s=0 states belonging to N$_e$=2 and N$_e$=4 respectively
one has to expect a complex magnetic behaviour of the model. For example a small
magnetic field lifts the degeneracy of the N$_e$=3 states whereby the states 
No 46 and 50 are lowered below the states No. 22 and No. 111. This process is shown
in Fig. \ref{spectrumU4N3h0low}. By further increasing the correlation energy U
we found a spin change of the groundstate, since the S=3/2 states no. 38, 58, 82, 
and 90 become the groundstates. This was first discovered by Callaway et.al. 
\cite{Callaway87} via numerical diagonalization. We add that the groundstate is 
fourfold degenerated, and the exact U value, where the transition happens is
U/t=4(2+$\sqrt{7}$), what differs slightly from the numerical value 1/0.055
given in Ref.~\onlinecite{Callaway87}. The difference is in the second digit and may serve
as an estimation of the accuracy of their calculation. For N$_e$=5
they did not find such a change in the groundstate spin, what is contradictory
to the electron-hole symmetry, which was otherwise explicitly mentioned in their 
paper - and of course: the same transition happens at the same U value.  
\subsection{The contents of double occupied states}
A further question of interest is whether projecting out the double-occupied
states from beginning is a good approximation. 
Since we have all the 
eigenstates in analytical form, it was easy to calculate the contents 
of double occupied states in each state. The U-dependence of the admixing of 
double occupied states to the lowest lying states is depicted 
in Fig. \ref{DoubleOccupied}.
Among the considered states the groundstate has the highest content for 
bigger U, being about ten per cent for U=10t and 
falling to about two per cent for U=30t. If the correlation
energy U is lower than the bandwidth the double occupied states admix
considerably. Thus one has to be careful with the projection, especially if
one aims at a description of the Hubbard model in the region were correlation
energy and bandwidth are of the same order.          
\subsection{The spectral function}
\label{spectralfun}
By help of the eigenstates and the spectrum it is possible to calculate the
spectral function for arbitrary operators. 
Since the $\delta$-like peaks 
result in discontinuous curves being inconvenient for the graphics 
we substituted the $\delta$-functions by Lorentz distributions, i.e. we
defined
\beq
J_{BA}(\omega)&=&\frac{2\pi}{Z}\sum_{m}\sum_{n} e^{-\beta E_m}  
\bra{\Phi_m}\operator{B}\ket{\Phi_n}\bra{\Phi_n}\operator{A}\ket{\Phi_m} \times 
\nonumber \\
&& \hspace{2cm} \times \frac{1}{\pi}\frac{\delta}{\delta^2+(\omega-E_m+E_n)^2}
\eeq
with Z being the grand canonical partition sum, and $E_m$ being the grand
canonical levels. The most interesting
function due to its relation to spectroscopy we get with 
$\operator{B}=\cminus{i\sigma}$ and $\operator{A}=\cplus{i\sigma}$.
In Fig. \ref{fig:6} we show the results in dependence of U for the case of
half-filling.
The most ocular feature of the spectral function is the
opening of the gap of width U with increasing correlation strength. This
is in qualitative accordance with the common picture of the 
Hubbard-Mott transition.
\section{Thermodynamics}
Once calculated the spectrum in an analytical way it is straight forward
to get the grand canonical potential and the complete thermodynamics.
For the four point cluster there exist several papers, 
which got the thermodynamics by numerical diagonalization. Heinig
and Monecke calculated for this model the specific heat \cite{Heinig72a}, 
the susceptibility and local magnetic moments \cite{Heinig72b} for the
half filled case. Shiba and Pincus did the same within a comprehensive work
devoted to the one dimensional half-filled-band Hubbard model\cite{Shiba72}. 
The discrepancies of these two works are discussed in a paper of Cabib and Kaplan,
who calculated in contrast both the canonical and 
the grand canonical ensemble \cite{Cabib73}. They especially discussed the
specific heat and the spin-spin correlation functions, where the main focus
was again the half-filled case. In the following we give our results for
both the half-filled case (for comparison) and the doped case.
\subsection{Thermodynamic density of states}  
In their paper \cite{Cabib73} Cabib and Kaplan 
argued that the grand canonical ensemble for a finite N system reflects the 
infinite N case much better than the canonical ensemble, especially for the
large U case. We agree with
that point of view, especially regarding 
the thermodynamical density of states (TDOS).   
We calculated this quantity, i.e.
the derivation of the particle number with respect to the chemical potential.
\beq
D_{T}(\mu)&=&\frac{ \partial N(\mu)}{\partial \mu}  \, .
\eeq
In Fig. \ref{fig:7} we depicted the results for different values of U at 
the relatively high temperature of $T=0.1t$ for graphical reasons.
It is obvious, that the TDOS changes its character qualitatively with
increasing U starting from a three-peak structure for small on site 
correlation changing to an eight-peak structure at large correlation.
Around $U=1t$ there opens a gap around $\mu=U/2$. This means, adding
a further electron to the half-filled model needs an energy of order U. 
We interpret this, in the same sense as above for the spectral function,
as the variant of the ``Hubbard-Mott-transition'' for our small model.
Of course, the ``critical U", where the gap opens, depends on
the temperature. For $T=0$ the gap opens for infinitesimal small values of
the correlation. This is a special case for the same statement, proved
for the one-dimensional Hubbard model by help of the exact solution for
the groundstate given by Lieb and Wu \cite{Lieb68}. 
\subsection{Specific heat}
The temperature dependence was controversial in
Refs.~\onlinecite{Heinig72a,Shiba72,Cabib73}.
For the half-filled case we find exactly the same features as given in
Ref.~\onlinecite{Cabib73}. Cabib et. al. stated the qualitative picture given in
Ref.~\onlinecite{Heinig72a} to be right and that of Ref.~\onlinecite{Shiba72} to be wrong for
the case U$<$6. Unfortunately a quantitative comparison with Ref.~\onlinecite{Heinig72a}
was impossible due to inconsistencies in that paper.
In Fig. \ref{fig:8}
we show the specific heat curves in dependence on the temperature. The curves
with U=0.7t and U=8t respectively are exactly the same as given in Fig. 1
of Ref.~\onlinecite{Cabib73} and thus  confirms this paper and serves as independent proof
of our calculation. Furthermore, the change from the double peak shape to
the triple peak structure is due to the behaviour of state 119, 
what is evident from Fig. \ref{fig:9}.
Variation of the electron density causes dramatic changes of the specific heat.
In Fig. \ref{fig:10} we calculated for a constant electron
density and constant magnetic field (for simplicity set to zero).
The shape of the specific heat curves is qualitatively affected if 
the electron density is changed.
In the right column of Fig. \ref{fig:10} the ratio of correlation energy U 
and band width is
chosen to 1. This is the most interesting case, since it is not accessible by
means of perturbation theory. We see that with deviation from half-filling
the small peak at k$_B$=0.1t vanishes, whereas a low energy peak develops
at k$_B$T=0.002t, being maximum at N$_e$=3, i.e. n$_e$=0.75. The reason is
the small distance between the states 46, 50, 70, 74 belonging  to N$_e$=3,
and the state 111, belonging to N$_e$=4, and state 22, belonging to N$_e$=2,
as may be seen from Fig. \ref{spectrumU5N3h01}. Since the N$_e$=3 
groundstate is magnetic and the groundstates for N$_e$=2 and N$_e$=4
respectively are not, 
these excitations are both spin excitations and charge excitations. In the right
column of Fig. \ref{fig:10} the correlation strength is twice the bandwidth.
In that case we find no small energy peak for N$_e$ deviating from 4. Instead
the double peak structure smears out due to the increasing influence of 
magnetic excitations which are maximum if the electron number is fixed to three.
Further decreasing the electron number results in vanishing of the double peak
structure. The reason is that the number of accessible states with double
occupation is reduced drastically.
\subsection{The crossing points}
A further subject of interest are the so called
crossing points, which attracted much interest  recently, since they are present
in several strong correlated systems. Vollhardt
clarified this point \cite{Vollhardt97}. In a second paper \cite{Chandra98} 
he and the coworkers specified to Hubbard models and found a (nearly) universal
crossing point of the C(T,U)/k$_B$-curves. Since the arguments in
Ref.~\onlinecite{Vollhardt97} never use the thermodynamic limit, we feel that they should
work for our small model also. In Fig. \ref{fig:11}
we depicted the specific heat curves nearly in the same fashion as was done in
Fig. 1 (b) of Ref.~\onlinecite{Chandra98}. The similarity is ocularly. The most striking
point the authors made in Ref.~\onlinecite{Chandra98} is the discovery 
of a nearly universal value of about 0.34 of the specific heat at the high
temperature crossing point for small U. We find for our model a value of 
the crossing temperature T$^{\star}$(U$\to$0)/t=1.334820204t and 
C(U$\to$0,T$^{\star}$)=0.3351877115 Nk$_B$. 
All the Hubbard systems considered in Ref.~\onlinecite{Chandra98} are in the 
thermodynamic limit at half-filling.  
Whether our small cluster is comparable to infinite systems is a
difficult question. Usually the physical quantities calculated
from the canonical and grand-canonical potentials differ in small systems. 
Therefore 
this difference may serve as a rough indicator how relevant the results are for
the thermodynamic limit. In Fig. \ref{fig:12} we plotted the specific heat for
the canonical ensemble. 
We found again a high temperature crossing point, whereas
there is no sharp crossing point around k$_B$T/t=0.8, in contrast to
Fig. \ref{fig:11}. The related specific heat value is 
C(U$\to$0,T$^{\star}$)=0.3744085787 Nk$_B$ with 
T$^{\star}$=0.940347332t. This is only  
a little bit bigger than the grand-canonical value.   
Nevertheless the big difference in T$^{\star}$ shows that there are
differences at all. The crossing points at different U values 
were determined according to Ref.~\onlinecite{Vollhardt97} by help of the equation
\beq
\frac{\partial C(T,U,\mu)}{\partial U} &=& 0 \qquad \mbox{with}\quad \mu=U/2 
\eeq
for the grand canonical ensemble and by
\beq
\frac{\partial C(T,U)}{\partial U} &=& 0 
\eeq
for the canonical ensemble with N$_e$=4 respectively. 
These functions significantly differ from each other, 
as one can see from Fig. \ref{fig:13}.
\subsection{Susceptibility}
The isothermal susceptibility shows the interplay between the low-lying 
nonmagnetic states and the low-lying magnetic states. 
As shown above, the fact whether the 
groundstate is magnetic or not is dependent on the occupation. 
That is why we find an increasing
peak due to the low-energy
magnetic excitations if we decrease the electron number from four to three, 
which vanishes, if we subsequently decrease the number of electrons towards 
two.
This is illustrated in Fig. \ref{fig:14}.
In the strong correlated case (U/t=8) we find a complete different
behaviour. Now the distance between the states with different electron
occupation is large. Consequently, we do not find any low-energy peaks as
visible in Fig. \ref{fig:15}. Instead we see a divergent behaviour due
to the S$_z$ degenerated groundstate if the electron number deviates from
four. 
It is well believed, that the Hubbard model in three dimensions exhibits 
antiferromagnetism near the half-filled limit and for low temperatures.
In our small model long range order does not exist, nevertheless we can
ask how the correlation functions depend upon change of the correlation 
strength, of temperature and electron filling.
We adopt the notation of Cabib and Kaplan \cite{Cabib73} for the
spin-spin correlation functions
\beq
L_n&=&4\expect{\operator{S}_i^z\operator{S}_{i+n}^z}
\eeq
If not otherwise mentioned we always take the expectation with respect to the
grand canonical ensemble. By help of the functions $L_n$ one can calculate
the susceptibility for our small cluster according to
\beq
\chi^T&=& 4 \beta (L_0 + 2 L_1 + L_2) - 4\beta\expect{\operator{S}^z} \,.
\label{chi2}
\eeq
The susceptibility calculated from eq. \ref{chi2} has to agree with the 
expression we got from the differentiation of the thermodynamical potential.
We used this fact to check the calculated correlation functions.
In Fig. \ref{fig:16} the three different spin-spin-correlation functions are
shown in dependence on the temperature for the half-filled case. The model 
shows the well known antiferromagnetic coupling of adjacent spins. Of course
this breaks down if the temperature is increased.The breakdown shifts to lower
temperatures when the strong coupling regime is reached. This is exactly what
is expected, due to the fact that for large values of U the model can be mapped
onto an antiferromagnetic Heisenberg model with the coupling constant being
$\propto$ 1/U. Obviously the U=4t case and the U=8t case have nearly the same
breakdown temperature, fortifying the above stated fact, that U=8t, i.e. U is
twice the unperturbed bandwidth, belongs not to the strong coupling limit. 
It is exactly this parameter regime where we find an increase of the local 
magnetic moment with increasing temperature before it breaks down to its high
temperature limit. The strong correlation behaviour of the model is shown in Fig.
\ref{fig:17}.  
The most striking fact is the change  from antiferromagnetic correlation to
ferromagnetic correlation if the electron density is lowered from one to
three quarters. Also the next nearest neighbour correlation changes its
sign when the electron occupation changes from three towards two. While
for the neutral model the nearest neighbours are antiferromagnetically correlated,
the next nearest neighbours are ferromagnetic. For three electrons both
correlation functions are ferromagnetic and for two electrons they are both
antiferromagnetic. 
The temperature dependence of the density-density 
correlations is depicted for the neutral model in Fig. \ref{fig:18}.
Regarding $\langle \operator{n}_1\operator{n}_1 \rangle=\langle
\operator{n}_1+2\operator{d}\rangle$ 
with $\operator{d}$ being the double occupancy operator 
$\operator{d}=\operator{n}_{\downarrow} \operator{n}_{\uparrow}$ it is obvious 
how the onsite electron correlation suppresses the probability to find 
two electrons at the same lattice site as long, as the correlation energy is
small with respect to the temperature. Especially the correlation function
$\langle \operator{n}_1\operator{n}_2 \rangle$ makes evident that the tendency
to charge ordering is lowered. We remark, that the TDOS may
be calculated via the density-density correlation functions, giving one more
possibility to prove the inner consistency of the derived formulae.
In Fig. \ref{fig:19} we depicted the nearest-neighbour density correlation for 
strong correlation in dependence of the electron filling of the model.
With decreasing electron number the electrons get more space. Due to the
repulsive electron-electron correlation the electrons tend to increase their
distance. Therefore the nearest-neighbour density-density correlation is
reduced considerably.
\section{Discussion}
The benefit one gets from an analytical solution of a model is the easy access
to its complete physics. Due to the limited space, we had to restrict the
presentation of the results to a few points. We focused here mainly to 
two points. The first one was the study of eigenvalues and related
eigenstates in dependence on the correlation strength. This way we clarified 
the open questions about possibly hidden symmetries. It was shown, that
every state is distinguished from each other by using the two SU[2] symmetries
of spin and pseudospin, the translation symmetry, and the eigenvalues of
the Grosse operator $\operator{K}_1$. Since therefore all the eigenstates
are off different symmetries, there remains no problem with the multitude
of level crossings present within the model. The lowest states in the
half-filled model yield the singlet-triplet picture, widely used 
in the context of high-T$_c$ superconductivity. In the
half-filled model the charge excitations are a little bit higher in energy, 
but, these excitations are lowered considerably in energy, if 
the electron number is driven towards three by appropriate tuning of the
chemical potential, thus enlightening the complex interplay of
spin and charge (more exactly: pseudospin) degrees of freedom. 
We were really surprised that the states with adjacent electron numbers are
still lower in energy if three is chosen for the mean occupation, even if
the correlation equals the bandwidth. This situation is completely changed if
one applies a (relatively) small magnetic field or if the ratio U/t
is increased slightly ( what may be done by help of pressure in real systems).
Since our model is a special case of the n-site Hubbard rings which are the pet
of the Bethe ansatzers, there must be a correspondence between the Bethe 
ansatz eigenstates
and the eigenstates given in the Appendix. However, it is not obvious,
how the quantum numbers characterizing the states in our
solution correspond to the quantum numbers of the theories based on the
Bethe ansatz. Two of them, the z-component of spin and pseudospin respectively
are used within the Bethe ansatz, but regarding the other quantum numbers 
of the Bethe ansatz, i.e the charge momenta and the spin
rapidities, we have no idea how they are related to the remaining quantum numbers
of our model, i.e. the modules of spin and pseudospin, the eigenvalue of the 
translation operator, and the eigenvalue of the Grosse operator. 
The complicated numerical solution of the Lieb-Wu equations 
delivering the charge momenta 
and spin rapidities prevents a deeper insight, so it remains an open question.
The interesting problem whether the eigenstates may be classified 
as k strings and k-$\lambda$ strings as Deguchi et. al. \cite{Deguchi99} did 
with respect to solutions of the Lieb-Wu equations following 
the string hypothesis of Takahashi \cite{Takahashi72}, or in the way as was 
done in a very recent paper \cite{Li00} remains a unsolved question for the time
being. 
A second point we were interested in, is whether the model gives some hints
towards the metal-insulator transition. Both the spectral density and
the thermodynamical density of states show the expected behaviour, i.e.
the opening of a gap of width U for T$>$0,
so there is nothing spectacular in this direction. The general theorems
for the Hubbard model are fulfilled of course.
Our analytical solution allowed a detailed study of the crossing points 
of the specific heat. Our results confirm the explanation given
by Vollhardt et. al. completely. At the first glance this seems astonishing, 
since we
have a level spectrum instead of the various continuous DOS' used in
Ref.~\onlinecite{Chandra98}. On the other hand, if one deals with the high-temperature 
crossing point the temperature smoothes the level structure in some sense. 
Furthermore, the argument, given in Ref.~\onlinecite{Chandra98}, that the energy scale 
for T is essentially t if we are at higher temperatures and of the order 
of the singlet-triplet excitation at lower temperatures holds for our model 
also. The comparison of the canonical and grand canonical ensemble at half
filling exhibit differences regarding the temperature dependence of the
derivatives of specific heat with respect to the correlation strength.The reason 
is that a ``sampling'' of the DOS is much poorer with 72 states instead of 256.
Thus, it is not unexpected that the values of C$^+$ differ slightly from the
value calculated in Ref.~\onlinecite{Chandra98} for an infinite linear chain by help of
second order perturbation theory in U/t. 
Considering the specific heat for deviations of the mean occupation number 
from half-filling 
we find another much lower energy scale, due to the fact that states with
different electron numbers are nearly degenerated if we fix the mean electron
number to three.  
Regarding the magnetic behaviour we see the interplay between the even-occupied
clusters, 
which are nonmagnetic and exhibit low-lying singlet-triplet
excitations and the odd occupied clusters which exhibit a magnetic degenerated
groundstate with doublet-quartet excitations if we have N$_e$=3 or N$_e$=5. 
By fixing the mean occupation number to non-integers we enforce a mixture of
both cluster groups yielding a diverging susceptibility due to the contents of
magnetic clusters.
The spin-spin correlations  confirm
the common accepted picture, that the strong on-site correlation favours 
antiferromagnetism instead ferromagnetism. Otherwise, next-nearest
neighbours are coupled ferromagnetic. This holds for the half filled model. 
This situation is changed completely if the mean occupation number is driven
towards three. There we find ferromagnetic coupling between adjacent spins
for low temperature. A completely unexpected behaviour is the
change to an antiferromagnetic coupling (albeit small) when the temperature
is increased and becomes of the order of the hopping matrix element.
For higher temperatures the off-site spin-spin correlation vanishes.
Since we have a small cluster long range magnetic order is impossible, 
nevertheless, if one regards the nearest neighbour correlation as an hint to
the
magnetic phases realized in an infinite system, our findings support the 
common accepted phase diagram for the (simple cubic) three-dimensional Hubbard model.
The suppression of electron motion with increasing correlation strength
is evident from the density-density-correlations. If U is large enough we find
exact one electron at every site. Motion seems forbidden, what hints towards 
an insulator.
Regarding the square lattice we believe that it is more
appropriate, with reference to the conserved quantities of the Hubbard model,
to speak about spin-pseudospin separation instead of spin-charge separation,
since the ``charge'' is the ``z-component'' of the pseudospin only. From models
in magnetism it is common knowledge that quantum fluctuations reduce
the magnetic moment, due to the fact that the total spin is conserved, while
the local spin is usually not. In the Hubbard model the same holds for the
pseudospin also. If the hopping between the clusters is treated by a
perturbation theory as was done in Ref.~\onlinecite{Senechal00}, the calculated cluster
levels will form bands, and the related cluster-spins and -pseudospins will not
be conserved furthermore, thus ``pseudospin-waves'' will arise as well as
spin-waves. The undoped model has a vanishing total pseudospin.
Reducing the number
of electrons from one per lattice cite creates a certain number of empty
places. At these sites a local spin is missing now and a local
pseudospin is created. Further doping dilutes the spins and increases the distance, what reduces
the spin-spin interaction. Otherwise the number of pseudospins increases; their 
mean distance is reduced and the pseudospin-pseudospin interaction is
increased, thus favouring long-range order of the pseudospins. 
As we have shown, among the lowest eigenstates of our cluster we find states
with different spins and pseudospins. Since the related states are many-particle 
states the spins and the pseudospins are not coupled to individual electrons or
holes. If these many-particle states become itinerant by switching on the
inter-cluster-hopping, there is no reason, that the resulting bands must exhibit
identical shapes, or in  other words that the resulting quasiparticles have the
same mass. On the contrary, one would expect, that a quasiparticle made from a
cluster-state with three electrons and spin 1/2 will differ significantly from 
a quasiparticle made from a spinless cluster-state with four (or two)
electrons, both in the effective mass and in the velocity of propagation. This
is exactly, what was calculated in Ref.~\onlinecite{Senechal00} and what is called
spin-charge separation.
Finally, we mention that the given solution is
applicable to the four-site Hubbard model with attractive interaction
as well, simply by applying the Shiba map \cite{Shiba72} 
$\cpup{i} \rightarrow {\bf a}^+_{i\uparrow}$ and $\cdown{i} \rightarrow {\bf a}^+_{i\downarrow}$, 
which exchanges the spin with the pseudospin. The tables given in the 
appendix apply as well for negative U hamiltonian. The only task is to
replace $ U \rightarrow -U$, $h+U/2 \rightarrow \mu$,
and $\mu+U/2 \rightarrow h$ for the energy eigenvalues and to exchange
the spin and pseudospin quantum numbers in the assignment of the
eigenvectors. All the presented figures, may be calculated for the attractive
model as well.    
There is a multitude of 
other interesting features, 
e.g. the transport properties within the cluster, 
which are nevertheless easy to
calculate by help of the analytically given eigenstates and eigenvectors.
We abstain from presenting the related figures 
to prevent overloading of the present paper.
In summary we may say that the analytical solution of the four-site Hubbard
model for arbitrary electron filling, arbitrary interaction strength and
arbitrary temperature is highly non-trivial. It allows easy access to all
interesting features of the model and may serve as a reference object
for various numerical or perturbation methods dealing with more complex models
of strong electron correlation.
\acknowledgements
I would like to thank S.-L. Drechsler and R. Hayn for
usefull discussion, J. Monecke for some helpful information on their 
early work, and P. Vollhardt for encouraging remarks on the crossing point
calculation.
\appendix
\section*{Eigenvectors and Eigenvalues} 
Here we present some eigenstates and the complete spectrum. Since by far the
most papers deal with the groundstate of the Hubbard model we restrict to
presentation of the groundstate of the neutral model with N$_e$=4, i.e. state 111, and one
of the groundstates for N$_e=3$, i.e. state 46. The analytical form of the
other 254 eigenvectors the interested reader may find at the web-page
www.physik.tu-dresden.de/itp/members/schumann/research.html.
\beq
\Psi_{111}&=& C_{1} \left ( \ket{0022} - \ket{0220} - \ket{2002} + \ket{2200}\right) \nonumber \\
&& + C_{2} \left ( - \ket{02du} + \ket{02ud} + \ket{0du2} - \ket{0ud2} \right)\nonumber \\
&& + C_{2} \left ( - \ket{20du} + \ket{20ud} + \ket{2du0} - \ket{2ud0} \right) \nonumber \\
&& + C_{2} \left ( + \ket{d02u} + \ket{d20u} - \ket{du02} - \ket{du20} \right) \nonumber \\ 
&& + C_{2} \left ( - \ket{u02d} - \ket{u20d} + \ket{ud02} + \ket{ud20} \right) \nonumber \\
&& + C_{3} \left ( \ket{dudu} + \ket{udud}\right) \nonumber \\
&& + C_{4} \left ( \ket{dduu} + \ket{duud} + \ket{uddu} + \ket{uudd}\right) \\ 
C_{1}&=&\frac{1}{2\,{\sqrt{3}}} + \frac{4\,\imag
\,{\sqrt{3}}\,U\,Y^{\frac{1}{3}}}{N_1} \nonumber  \\
C_{1}&=& 96\,2^{\frac{1}{3}}\,(\sqrt{3}-\imag )\,t^2 
       +  6\,2^{\frac{1}{3}}\,({\sqrt{3}}-\imag)\,U^2 \nonumber \\
   &&  - 12\,\imag \,U\,Y^{\frac{1}{3}} 
       - 2^{\frac{2}{3}}\,({\sqrt{3}}+\imag)\,Y^{\frac{2}{3}} \nonumber \\
C_{2}&=&  \frac{-U}{8\,{\sqrt{3}}\,t} 
        + \frac{(1+\imag\sqrt{3})\,2^{\frac{1}{3}}\,t}{\sqrt{3}\,Y^{\frac{1}{3}}} \nonumber \\ 
     && + \frac{(1+\imag\sqrt{3})\,U^2}{8\,2^{\frac{2}{3}}\,{\sqrt{3}}\,t\,Y^{\frac{1}{3}}} 
        + \frac{(1-\imag\sqrt{3})\,Y^{\frac{1}{3}}}{48\,2^{\frac{1}{3}}\,{\sqrt{3}}\,t}\nonumber \\
C_{3}&=&-1 + \frac{1}{{\sqrt{3}}}\nonumber \\
C_{4}&=&\frac{1}{2} + \frac{1}{{\sqrt{3}}}\nonumber \\
Y&=&-216\,t^2\,U + 6\,{\sqrt{3}}\,X\nonumber  \\
X&=&{\sqrt{-4096\,t^6 - 336\,t^4\,U^2 - 48\,t^2\,U^4 - U^6}}\nonumber 
\eeq
\beq
\Psi_{46}&=& C_{1} \left ( -\imag \,\ket{0duu} + \ket{duu0} - \ket{u0du} + \imag \,\ket{uu0d}\right) \nonumber \\
&& + C_{2} \left ( -\ket{0uud} + \imag \,\ket{d0uu} + \ket{ud0u} - \imag \,\ket{uud0}\right) \nonumber \\
&& + C_{3} \left ( -\imag \,\ket{0udu} + \imag \,\ket{du0u} - \ket{u0ud} + \ket{udu0}\right) \nonumber \\
&& + C_{4} \left ( \imag \,\ket{002u} - \ket{02u0} - \imag \,\ket{2u00} + \ket{u002}\right) \nonumber \\
&& + C_{5} \left ( \imag \,\ket{020u} - \imag \,\ket{0u02} - \ket{20u0} + \ket{u020}\right) \nonumber \\
&& + C_{6} \left ( \ket{00u2} + \imag \,\ket{0u20} - \imag \,\ket{200u} - \ket{u200}\right) \\
C_{1}&=&   \frac{1}{2\,{\sqrt{2}}} 
         + \frac{\frac{3\,\imag }{4}\,U}{{\sqrt{2}}\,t}
         - \frac{X}{4\,{\sqrt{2}}\,t^2} 
         + \frac{\frac{3\,\imag}{4}\,Y}{{\sqrt{2}}\,t}\nonumber \\
     &&  + \frac{8\,\imag \,{\sqrt{2}}\,t\,Y}{U^2} 
         - \frac{   \imag \,{\sqrt{2}}\,X\,Y}{t\,U^2}
         + \frac{6\,\imag \,{\sqrt{2}}\,t}{U} \nonumber \\
     &&  - \frac{\frac{3\,\imag }{2}\,X}{{\sqrt{2}}\,t\,U} 
         + \frac{{\sqrt{2}}\,Y}{U} 
         - \frac{X\,Y}{4\,{\sqrt{2}}\,t^2\,U} \nonumber \\
C_{2}&=&   \frac{-\imag}{3\,{\sqrt{2}}} 
         - \frac{6\,{\sqrt{2}}\,t}{U} 
         - \frac{3\,U}{4\,{\sqrt{2}}\,t} 
         + \frac{\imag \,X}{6\,{\sqrt{2}}\,t^2} \nonumber \\
      && + \frac{3\,X}{2\,{\sqrt{2}}\,t\,U} 
         - \frac{Y}{2\,{\sqrt{2}}\,t} 
         - \frac{\imag \,{\sqrt{2}}\,Y}{U} \nonumber \\
      && + \frac{\imag \,X\,Y}{4\,\sqrt{2}\,t^2\,U}
         + \frac{-8\,{\sqrt{2}}\,t\,Y}{U^2} 
         + \frac{{\sqrt{2}}\,X\,Y}{t\,U^2} \nonumber \\
C_{3}&=&   \frac{-1}{3\,{\sqrt{2}}} 
         + \frac{X}{6\,{\sqrt{2}}\,t^2} 
         - \frac{2\,{\sqrt{2}}\,Y}{3\,U} 
         + \frac{X\,Y}{6\,{\sqrt{2}}\,t^2\,U}\nonumber \\
C_{4}&=&   \frac{-\imag }{{\sqrt{2}}} 
         - \frac{4\,{\sqrt{2}}\,t}{3\,U} 
         + \frac{X}{3\,{\sqrt{2}}\,t\,U} 
         - \frac{8\,{\sqrt{2}}\,t\,Y}{3\,U^2} 
         + \frac{{\sqrt{2}}\,X\,Y}{3\,t\,U^2}\nonumber \\
C_{5}&=&   \frac{-4\,{\sqrt{2}}\,t}{3\,U} 
         + \frac{X}{3\,{\sqrt{2}}\,t\,U} 
         - \frac{8\,{\sqrt{2}}\,t\,Y}{3\,U^2} 
         + \frac{{\sqrt{2}}\,X\,Y}{3\,t\,U^2}\nonumber \\
C_{6}&=&   \frac{-\imag }{{\sqrt{2}}} 
         + \frac{\sqrt{2}}{3}\, \left(
           \frac{4\,t}{U} 
         - \frac{X}{2\,t\,U} 
         + \frac{8\,t\,Y}{U^2} 
         - \frac{X\,Y}{t\,U^2}\right)\nonumber \\
X&=&{\sqrt{64\,t^4 + 3\,t^2\,U^2}}\nonumber  \\
Y&=&\sqrt{32\,t^2 + U^2 + 4\,X}   \nonumber 
\eeq
In following Tables \ref{table:n1}-\ref{table:n8} we give the results for the whole spectrum. The
eigenvalues of the operators 
$\rtot_z$ ,$ \rtot$,$ \stot_z$,$ \stot^2$ ,$ \operator{U}$,$\operator{K}_{11}$,
$\operator{P}^{++}$,$ \operator{P}^{+-}$, $\hamilton$ are assigned as 
$m_r$, $m_s$, $r$, $s$, $u$, $k_{11}$, $p_{++}$, $p_{+-}$, $\varepsilon$. 
In the first column we number the eigenstates. 
In the second column the energy eigenstates are given in the form
\tket{m_r, m_s,r(r+1),s(s+1), u, k_{11}, p_{++}, p{+-}}. 
The related energy eigenvalues in dependence on the parameters t and U are 
given in the third column. 
For the grand canonical eigenvalues these
energy eigenvalues have to be completed by $-\mu {N_e} - h m_s$, therefore we 
give the related values in the subheads of the tables. 
In the fourth column we calculated the numbers for the energy for t=1, U=5, 
$\mu=U/2$, and $h=0.01$, i.e. the half filled case. The small magnetic field
was introduced to lift spin-degeneracy. \\

%
\newpage
\begin{figure}
\caption{\label{abb: clustergas} By replacing every second hopping line by an
indirect
hopping via the bath the square-lattice (A) is transformed to the
square-cluster gas (B).}
\label{clustergas}
\end{figure}
\begin{figure}
\caption{\label{abb:spectrumU5N4h01} The low lying states of the half-filled
model. The parameters are t=1, U=5, h=0.01, \texpect{N_e}=4. The main quantum numbers
are indicated at the levels. The length indicates the degeneracies.}
\label{spectrumU5N4h01}
\end{figure}
\begin{figure}
\caption{\label{abb: NviaMueU5h01} The particle number in dependence
on the chemical potential. The parameters are t=1, U=5, h=0.01.}
\label{NviaMueU5h01}
\end{figure}
\begin{figure}
\caption{\label{abb: GSviaMueU5h01} The grand canonical groundstate in dependence on
the chemical potential. The parameters are t=1, U=5, h=0.01 .}
\label{GSviaMueU5h01}
\end{figure}
\begin{figure}
\caption{\label{abb: spectrumU5N3h01} The low lying states for \tmean{N_e}=3. 
The parameters are t=1, U=5, h=0.01 (h=0 for the dotted 
curve), k$_{B}$T=10$^{-7}$.}
\label{spectrumU5N3h01}
\end{figure}
\begin{figure}
\caption{\label{abb: spectrumU4U5N3h0low} The lowest states for \tmean{N_e}=3
and U=4t and U=5t respectively. 
The remaining parameters are t=1, h=0}
\label{spectrumU4U5N3h0low}
\end{figure}
\begin{figure}
\caption{\label{abb: spectrumU4N3h0low} The lowest states for \tmean{N_e}=3
and h=0t and h=0.01t respectively. 
The remaining parameters are t=1, U=4t}
\label{spectrumU4N3h0low}
\end{figure}
\begin{figure}
\caption{\label{abb: figure5} Contents of double-occupied states.
The parameters are t=1, h=0, \tmean{N_e}=4. The operator $\operator{P}$ projects out the
double-occupied states.}
\label{DoubleOccupied}
\end{figure}
\begin{figure}
\caption{\label{abb: figure6} The spectral function 
$J_{\cminus{i\sigma}\cplus{i\sigma}}$.
The parameters are t=1, h=0, \tmean{N_e}=4.}
\label{fig:6}
\end{figure}
\begin{figure}
\caption{\label{abb: figure7} The thermodynamical density of states.
The parameters are t=1, h=0, k$_{B}$T=0.1t.}
\label{fig:7}
\end{figure}
\begin{figure}[hbt]
\caption{\label{abb: figure8} The specific heat in dependence on the
temperature for different values of the correlation strength calculated from
the grand canonical partition sum.
The other parameters are t=1, h=0, \tmean{N_e}=4.}
\label{fig:8}
\end{figure}
\begin{figure}
\caption{\label{abb: figure9} The U-dependence of the low-lying levels
for the half-filled model.}
\label{fig:9}
\end{figure}
\begin{figure}
\caption{\label{abb: figure10} The specific heat in dependence on the
temperature for different electron filling. The number inside gives
the number of electrons.The parameters are t=1, h=0, and U=4 for the left
column and U=8 for the right column respectively.}
\label{fig:10}
\end{figure}
\begin{figure}
\caption{\label{abb: figure11} The  specific heat in dependence 
on the temperature for the grand-canonical ensemble at half-filling.
The upper picture shows at a large scale the crossing points, whereas the
inset shows the high temperature crossing point in fine scale. The
lower plot demonstrates that there is no definite high-temperature crossing 
point for large U values. The parameters are t=1, h=0, and the U values are
indicated at the curves.}
\label{fig:11}
\end{figure}
\begin{figure}
\caption{\label{abb: figure12} The specific heat in dependence on the
temperature for the canonical ensemble and half-filling.
The inset shows the high-temperature crossing point in fine scale. The
U values are the same as in fig. \ref{fig:11}.}
\label{fig:12}
\end{figure}
\begin{figure}
\caption{\label{abb: figure13} The partial derivative of the specific
heat with respect to U for the canonical and grand-canonical ensemble
respectively.}
\label{fig:13}
\end{figure}
\begin{figure}
\caption{\label{abb: figure14} The temperature dependence of the 
isothermal susceptibility for t=1 and U=4. The parameters at the curves 
indicate the electron number.}
\label{fig:14}
\end{figure}
\begin{figure}
\caption{\label{abb: figure15} The temperature dependence of the 
isothermal susceptibility for t=1 and U=8. The parameters at the curves 
indicate the electron number.}
\label{fig:15}
\end{figure}
\begin{figure}
\caption{\label{abb: figure16} The Spin-Spin-Correlations in dependence on the
temperature for the correlation parameters t=1, U=1, U=4, and U=8.
The parameters at the curves indicate the electron number.}
\label{fig:16}
\end{figure}
\begin{figure}
\caption{\label{abb: figure17} The nearest neighbour spin-spin-correlation 
function
 $\langle \operator{S}_1\operator{S}_2\rangle$ for strong correlation
U=40 in dependence on the
temperature. The parameters at the curves 
indicate the electron number.The inset shows the density dependence of the
spin-spin correlation in the groundstate. }
\label{fig:17}
\end{figure}
\begin{figure}
\caption{\label{abb: figure18} The density-density-correlations in dependence
on the temperature for the half-filled model and for the correlation parameters
t=1, U=1, U=4, and U=8t.}
\label{fig:18}
\end{figure}
\begin{figure}
\caption{\label{abb: figure19} The nearest neighbour density-density-correlation 
function
 $\langle \operator{n}_1\operator{n}_2\rangle$ for strong correlation
U=40 in dependence on the
temperature. The parameters at the curves 
indicate the electron number.The inset shows the density dependence of the
nearest-neighbour density-density correlation in the groundstate. }
\label{fig:19}
\end{figure}
\newpage
\onecolumn
\renewcommand{\arraystretch}{1.5}
\begin{table}[b]
\caption{Eigenkets and eigenvalues for ${\rm N_e}$=0.}
\begin{tabular}{rlcc}
Number& Eigenstate&\multicolumn{2}{c}{Energy eigenvalue}\\\hline
\hline \multicolumn{4}{c}{\bf ${\rm N_e}$=0 m$_s$=$0$ r(r+1)=$6$ s(s+1)=$0$}\\ \hline 
1&$\ket{-2,0,6,0,1,0,4,0}$&$ 0$&0 \\
\end{tabular}
\label{table:n1}
\end{table}
\begin{table}
\caption{Eigenkets and eigenvalues for ${\rm N_e}$=1.}
\begin{tabular}{rlcc}
Number& Eigenstate&\multicolumn{2}{c}{Energy eigenvalue  \hspace{1cm} Value for
U/t=5, $\mu$=U/2, and h=0.01t}\\\hline
\hline \multicolumn{4}{c}{\bf ${\rm N_e}$=1 m$_s$=${1\over 2}$
r(r+1)=${{15}\over 4}$ s(s+1)=${3\over 4}$}\\ \hline 
2&$\ket{-{3\over 2},{1\over 2},{{15}\over 4},{3\over 4},-1,0,0,-4}$&$ -2\,t$&-4.51 \\
3&$\ket{-{3\over 2},{1\over 2},{{15}\over 4},{3\over 4},-i,2\,i,0,0}$&$ 0$&-2.51 \\
4&$\ket{-{3\over 2},{1\over 2},{{15}\over 4},{3\over 4},i,-2\,i,0,0}$&$ 0$&-2.51 \\
5&$\ket{-{3\over 2},{1\over 2},{{15}\over 4},{3\over 4},1,0,4,0}$&$ 2\,t$&-0.51 \\
\hline \multicolumn{4}{c}{\bf ${\rm N_e}$=1 m$_s$=$-{1\over 2}$ r(r+1)=${{15}\over 4}$ s(s+1)=${3\over 4}$}\\ \hline 
6&$\ket{-{3\over 2},-{1\over 2},{{15}\over 4},{3\over 4},-1,0,0,-4}$&$ -2\,t$&-4.49 \\
7&$\ket{-{3\over 2},-{1\over 2},{{15}\over 4},{3\over 4},-i,2\,i,0,0}$&$ 0$&-2.49 \\
8&$\ket{-{3\over 2},-{1\over 2},{{15}\over 4},{3\over 4},i,-2\,i,0,0}$&$ 0$&-2.49 \\
9&$\ket{-{3\over 2},-{1\over 2},{{15}\over 4},{3\over 4},1,0,4,0}$&$ 2\,t$&-0.49 \\
\end{tabular}
\end{table}
\clearpage
\begin{table}
\caption{Eigenkets and eigenvalues for ${\rm N_e}$=2. 
The abbreviation $\alpha$ is defined as $\alpha=\arccos({{{4\,{t^2}\,U}\over 3} - 
{{{U^3}}\over {27}}}/{{{\left( {{16\,{t^2}}\over 3} + 
{{{U^2}}\over 9} \right) }^{{3\over 2}}}}) $.}
\label{table:n2}
\begin{tabular}{rlcc}
Number& Eigenstate&\multicolumn{2}{c}{Energy eigenvalue  \hspace{1cm} Value for U/t=5, $\mu$=U/2, and h=0.01t}\\\hline
\hline \multicolumn{4}{c}{\bf ${\rm N_e}$=2 m$_s$=$1$ r(r+1)=$2$ s(s+1)=$2$}\\ \hline 
10&$\ket{-1,1,2,2,-1,0,0,-4}$&$ 0$&-5.02 \\
11&$\ket{-1,1,2,2,-i,-2\,i,0,0}$&$ -2\,t$&-7.02 \\
12&$\ket{-1,1,2,2,-i,2\,i,0,0}$&$ 2\,t$&-3.02 \\
13&$\ket{-1,1,2,2,i,-2\,i,0,0}$&$ 2\,t$&-3.02 \\
14&$\ket{-1,1,2,2,i,2\,i,0,0}$&$ -2\,t$&-7.02 \\
15&$\ket{-1,1,2,2,1,0,-4,0}$&$ 0$&-5.02 \\
\hline \multicolumn{4}{c}{\bf ${\rm N_e}$=2 m$_s$=$0$ r(r+1)=$2$ s(s+1)=$0$}\\ \hline 
16&$\ket{-1,0,2,0,-1,-2\,i\,{\sqrt{2}},0,0}$&$ 0$&-5. \\
17&$\ket{-1,0,2,0,-1,2\,i\,{\sqrt{2}},0,0}$&$ 0$&-5. \\
18&$\ket{-1,0,2,0,-i,0,0,0}$&$ {U\over 2} - {{{\sqrt{16\,{t^2} + {U^2}}}}\over 2}$&-5.70156 \\
19&$\ket{-1,0,2,0,-i,0,0,0}$&$ {U\over 2} + {{{\sqrt{16\,{t^2} + {U^2}}}}\over 2}$&0.701562 \\
20&$\ket{-1,0,2,0,i,0,0,0}$&$ {U\over 2} - {{{\sqrt{16\,{t^2} + {U^2}}}}\over 2}$&-5.70156 \\
21&$\ket{-1,0,2,0,i,0,0,0}$&$ {U\over 2} + {{{\sqrt{16\,{t^2} + {U^2}}}}\over 2}$&0.701562 \\
22&$\ket{-1,0,2,0,1,0,4,0}$&$ {U\over 3} - {{2\over 3}\,{\sqrt{48\,{t^2} +{U^2}}}\,
\cos ({{\alpha}\over 3})}$&-8.34789 \\
23&$\ket{-1,0,2,0,1,0,4,0}$&$ {U\over 3} + {{2\over 3}\,\,{\sqrt{48\,{t^2} +
{U^2}}}\,\cos ({{\pi  - \alpha}\over 3})}$&1.5136 \\
24&$\ket{-1,0,2,0,1,0,4,0}$&$ {U\over 3} + {{2\over 3}\,\,{\sqrt{48\,{t^2} +
{U^2}}}\,\cos ({{\pi  + \alpha}\over 3})}$&-3.16571 \\
&&$\alpha=\arccos \frac{{{4\,{t^2}\,U}\over 3} - 
{{{U^3}}\over {27}}}{{{\left( {{16\,{t^2}}\over 3} + 
{{{U^2}}\over 9} \right) }^{{3\over 2}}}}$&\\
\hline \multicolumn{4}{c}{\bf ${\rm N_e}$=2 m$_s$=$0$ r(r+1)=$2$ s(s+1)=$2$}\\ \hline 
25&$\ket{-1,0,2,2,-1,0,0,-4}$&$ 0$&-5. \\
26&$\ket{-1,0,2,2,-i,-2\,i,0,0}$&$ -2\,t$&-7. \\
27&$\ket{-1,0,2,2,-i,2\,i,0,0}$&$ 2\,t$&-3. \\
28&$\ket{-1,0,2,2,i,-2\,i,0,0}$&$ 2\,t$&-3. \\
29&$\ket{-1,0,2,2,i,2\,i,0,0}$&$ -2\,t$&-7. \\
30&$\ket{-1,0,2,2,1,0,-4,0}$&$ 0$&-5. \\
\hline \multicolumn{4}{c}{\bf ${\rm N_e}$=2 m$_s$=$0$ r(r+1)=$6$ s(s+1)=$0$}\\ \hline 
31&$\ket{-1,0,6,0,-1,0,0,-4}$&$ U$&0 \\
\hline \multicolumn{4}{c}{\bf ${\rm N_e}$=2 m$_s$=$-1$ r(r+1)=$2$ s(s+1)=$2$}\\ \hline 
32&$\ket{-1,-1,2,2,-1,0,0,-4}$&$ 0$&-4.98 \\
33&$\ket{-1,-1,2,2,-i,-2\,i,0,0}$&$ -2\,t$&-6.98 \\
34&$\ket{-1,-1,2,2,-i,2\,i,0,0}$&$ 2\,t$&-2.98 \\
35&$\ket{-1,-1,2,2,i,-2\,i,0,0}$&$ 2\,t$&-2.98 \\
36&$\ket{-1,-1,2,2,i,2\,i,0,0}$&$ -2\,t$&-6.98 \\
37&$\ket{-1,-1,2,2,1,0,-4,0}$&$ 0$&-4.98 \\
\end{tabular}
\end{table}
\begin{table}
\caption{Eigenkets and eigenvalues for ${\rm N_e}$=3, spin-up states}
\begin{tabular}{rlcc}
Number& Eigenstate&\multicolumn{2}{c}{Energy eigenvalue  \hspace{1cm} Value for U/t=5, $\mu$=U/2, and h=0.01t}\\\hline
\hline \multicolumn{4}{c}{\bf ${\rm N_e}$=3 m$_s$=${3\over 2}$ r(r+1)=${3\over 4}$ s(s+1)=${{15}\over 4}$}\\ \hline 
38&$\ket{-{1\over 2},{3\over 2},{3\over 4},{{15}\over 4},-1,0,0,4}$&$ -2\,t$&-9.53 \\
39&$\ket{-{1\over 2},{3\over 2},{3\over 4},{{15}\over 4},-i,-2\,i,0,0}$&$ 0$&-7.53 \\
40&$\ket{-{1\over 2},{3\over 2},{3\over 4},{{15}\over 4},i,2\,i,0,0}$&$ 0$&-7.53 \\
41&$\ket{-{1\over 2},{3\over 2},{3\over 4},{{15}\over 4},1,0,-4,0}$&$ 2\,t$&-5.53 \\
\hline \multicolumn{4}{c}{\bf ${\rm N_e}$=3 m$_s$=${1\over 2}$ r(r+1)=${3\over 4}$ s(s+1)=${3\over 4}$}\\ \hline 
42&$\ket{-{1\over 2},{1\over 2},{3\over 4},{3\over 4},-1,-i\,{\sqrt{3}},0,0}$&$ {U\over 2} - {{{\sqrt{16\,{t^2} - 4\,t\,U + {U^2}}}}\over 2}$&-7.30129 \\
43&$\ket{-{1\over 2},{1\over 2},{3\over 4},{3\over 4},-1,-i\,{\sqrt{3}},0,0}$&$ {U\over 2} + {{{\sqrt{16\,{t^2} - 4\,t\,U + {U^2}}}}\over 2}$&-2.71871 \\
44&$\ket{-{1\over 2},{1\over 2},{3\over 4},{3\over 4},-1,i\,{\sqrt{3}},0,0}$&$ {U\over 2} - {{{\sqrt{16\,{t^2} - 4\,t\,U + {U^2}}}}\over 2}$&-7.30129 \\
45&$\ket{-{1\over 2},{1\over 2},{3\over 4},{3\over 4},-1,i\,{\sqrt{3}},0,0}$&$ {U\over 2} + {{{\sqrt{16\,{t^2} - 4\,t\,U + {U^2}}}}\over 2}$&-2.71871 \\
46&$\ket{-{1\over 2},{1\over 2},{3\over 4},{3\over 4},-i,i,0,0}$&$ {U\over 2} - {{{\sqrt{32\,{t^2} + {U^2} + 4\,{\sqrt{64\,{t^4} + 3\,{t^2}\,{U^2}}}}}}\over 2}$&-10.1129 \\
47&$\ket{-{1\over 2},{1\over 2},{3\over 4},{3\over 4},-i,i,0,0}$&$ {U\over 2} + {{{\sqrt{32\,{t^2} + {U^2} + 4\,{\sqrt{64\,{t^4} + 3\,{t^2}\,{U^2}}}}}}\over 2}$&0.0929233 \\
48&$\ket{-{1\over 2},{1\over 2},{3\over 4},{3\over 4},-i,i,0,0}$&$ {U\over 2} - {{{\sqrt{32\,{t^2} + {U^2} - 4\,{\sqrt{64\,{t^4} + 3\,{t^2}\,{U^2}}}}}}\over 2}$&-6.57849 \\
49&$\ket{-{1\over 2},{1\over 2},{3\over 4},{3\over 4},-i,i,0,0}$&$ {U\over 2} + {{{\sqrt{32\,{t^2} + {U^2} - 4\,{\sqrt{64\,{t^4} + 3\,{t^2}\,{U^2}}}}}}\over 2}$&-3.44151 \\
50&$\ket{-{1\over 2},{1\over 2},{3\over 4},{3\over 4},i,-i,0,0}$&$ {U\over 2} - {{{\sqrt{32\,{t^2} + {U^2} + 4\,{\sqrt{64\,{t^4} + 3\,{t^2}\,{U^2}}}}}}\over 2}$&-10.1129 \\
51&$\ket{-{1\over 2},{1\over 2},{3\over 4},{3\over 4},i,-i,0,0}$&$ {U\over 2} + {{{\sqrt{32\,{t^2} + {U^2} + 4\,{\sqrt{64\,{t^4} + 3\,{t^2}\,{U^2}}}}}}\over 2}$&0.0929233 \\
52&$\ket{-{1\over 2},{1\over 2},{3\over 4},{3\over 4},i,-i,0,0}$&$ {U\over 2} - {{{\sqrt{32\,{t^2} + {U^2} - 4\,{\sqrt{64\,{t^4} + 3\,{t^2}\,{U^2}}}}}}\over 2}$&-6.57849 \\
53&$\ket{-{1\over 2},{1\over 2},{3\over 4},{3\over 4},i,-i,0,0}$&$ {U\over 2} + {{{\sqrt{32\,{t^2} + {U^2} - 4\,{\sqrt{64\,{t^4} + 3\,{t^2}\,{U^2}}}}}}\over 2}$&-3.44151 \\
54&$\ket{-{1\over 2},{1\over 2},{3\over 4},{3\over 4},1,-i\,{\sqrt{3}},0,0}$&$ {U\over 2} - {{{\sqrt{16\,{t^2} + 4\,t\,U + {U^2}}}}\over 2}$&-8.91512 \\
55&$\ket{-{1\over 2},{1\over 2},{3\over 4},{3\over 4},1,-i\,{\sqrt{3}},0,0}$&$ {U\over 2} + {{{\sqrt{16\,{t^2} + 4\,t\,U + {U^2}}}}\over 2}$&-1.10488 \\
56&$\ket{-{1\over 2},{1\over 2},{3\over 4},{3\over 4},1,i\,{\sqrt{3}},0,0}$&$ {U\over 2} - {{{\sqrt{16\,{t^2} + 4\,t\,U + {U^2}}}}\over 2}$&-8.91512 \\
57&$\ket{-{1\over 2},{1\over 2},{3\over 4},{3\over 4},1,i\,{\sqrt{3}},0,0}$&$ {U\over 2} + {{{\sqrt{16\,{t^2} + 4\,t\,U + {U^2}}}}\over 2}$&-1.10488 \\
\hline \multicolumn{4}{c}{\bf ${\rm N_e}$=3 m$_s$=${1\over 2}$ r(r+1)=${3\over 4}$ s(s+1)=${{15}\over 4}$}\\ \hline 
58&$\ket{-{1\over 2},{1\over 2},{3\over 4},{{15}\over 4},-1,0,0,4}$&$ -2\,t$&-9.51 \\
59&$\ket{-{1\over 2},{1\over 2},{3\over 4},{{15}\over 4},-i,-2\,i,0,0}$&$ 0$&-7.51 \\
60&$\ket{-{1\over 2},{1\over 2},{3\over 4},{{15}\over 4},i,2\,i,0,0}$&$ 0$&-7.51 \\
61&$\ket{-{1\over 2},{1\over 2},{3\over 4},{{15}\over 4},1,0,-4,0}$&$ 2\,t$&-5.51 \\
\hline \multicolumn{4}{c}{\bf ${\rm N_e}$=3 m$_s$=${1\over 2}$ r(r+1)=${{15}\over 4}$ s(s+1)=${3\over 4}$}\\ \hline 
62&$\ket{-{1\over 2},{1\over 2},{{15}\over 4},{3\over 4},-1,0,0,-4}$&$ 2\,t + U$&-0.51 \\
63&$\ket{-{1\over 2},{1\over 2},{{15}\over 4},{3\over 4},-i,-2\,i,0,0}$&$ U$&-2.51 \\
64&$\ket{-{1\over 2},{1\over 2},{{15}\over 4},{3\over 4},i,2\,i,0,0}$&$ U$&-2.51 \\
65&$\ket{-{1\over 2},{1\over 2},{{15}\over 4},{3\over 4},1,0,4,0}$&$ -2\,t + U$&-4.51 \\
\end{tabular}
\end{table}
\begin{table}\caption{Eigenkets and eigenvalues for ${\rm N_e}$=3, spin-down states.}
\begin{tabular}{rlcc}
Number& Eigenstate&\multicolumn{2}{c}{Energy eigenvalue  \hspace{1cm} Value for U/t=5, $\mu$=U/2, and h=0.01t}\\\hline
\hline \multicolumn{4}{c}{\bf ${\rm N_e}$=3 m$_s$=$-{1\over 2}$ r(r+1)=${3\over 4}$ s(s+1)=${3\over 4}$}\\ \hline 
66&$\ket{-{1\over 2},-{1\over 2},{3\over 4},{3\over 4},-1,-i\,{\sqrt{3}},0,0}$&$ {U\over 2} - {{{\sqrt{16\,{t^2} - 4\,t\,U + {U^2}}}}\over 2}$&-7.28129 \\
67&$\ket{-{1\over 2},-{1\over 2},{3\over 4},{3\over 4},-1,-i\,{\sqrt{3}},0,0}$&$ {U\over 2} + {{{\sqrt{16\,{t^2} - 4\,t\,U + {U^2}}}}\over 2}$&-2.69871 \\
68&$\ket{-{1\over 2},-{1\over 2},{3\over 4},{3\over 4},-1,i\,{\sqrt{3}},0,0}$&$ {U\over 2} - {{{\sqrt{16\,{t^2} - 4\,t\,U + {U^2}}}}\over 2}$&-7.28129 \\
69&$\ket{-{1\over 2},-{1\over 2},{3\over 4},{3\over 4},-1,i\,{\sqrt{3}},0,0}$&$ {U\over 2} + {{{\sqrt{16\,{t^2} - 4\,t\,U + {U^2}}}}\over 2}$&-2.69871 \\
70&$\ket{-{1\over 2},-{1\over 2},{3\over 4},{3\over 4},-i,i,0,0}$&$ {U\over 2} - {{{\sqrt{32\,{t^2} + {U^2} + 4\,{\sqrt{64\,{t^4} + 3\,{t^2}\,{U^2}}}}}}\over 2}$&-10.0929 \\
71&$\ket{-{1\over 2},-{1\over 2},{3\over 4},{3\over 4},-i,i,0,0}$&$ {U\over 2} + {{{\sqrt{32\,{t^2} + {U^2} + 4\,{\sqrt{64\,{t^4} + 3\,{t^2}\,{U^2}}}}}}\over 2}$&0.112923 \\
72&$\ket{-{1\over 2},-{1\over 2},{3\over 4},{3\over 4},-i,i,0,0}$&$ {U\over 2} - {{{\sqrt{32\,{t^2} + {U^2} - 4\,{\sqrt{64\,{t^4} + 3\,{t^2}\,{U^2}}}}}}\over 2}$&-6.55849 \\
73&$\ket{-{1\over 2},-{1\over 2},{3\over 4},{3\over 4},-i,i,0,0}$&$ {U\over 2} + {{{\sqrt{32\,{t^2} + {U^2} - 4\,{\sqrt{64\,{t^4} + 3\,{t^2}\,{U^2}}}}}}\over 2}$&-3.42151 \\
74&$\ket{-{1\over 2},-{1\over 2},{3\over 4},{3\over 4},i,-i,0,0}$&$ {U\over 2} - {{{\sqrt{32\,{t^2} + {U^2} + 4\,{\sqrt{64\,{t^4} + 3\,{t^2}\,{U^2}}}}}}\over 2}$&-10.0929 \\
75&$\ket{-{1\over 2},-{1\over 2},{3\over 4},{3\over 4},i,-i,0,0}$&$ {U\over 2} + {{{\sqrt{32\,{t^2} + {U^2} + 4\,{\sqrt{64\,{t^4} + 3\,{t^2}\,{U^2}}}}}}\over 2}$&0.112923 \\
76&$\ket{-{1\over 2},-{1\over 2},{3\over 4},{3\over 4},i,-i,0,0}$&$ {U\over 2} - {{{\sqrt{32\,{t^2} + {U^2} - 4\,{\sqrt{64\,{t^4} + 3\,{t^2}\,{U^2}}}}}}\over 2}$&-6.55849 \\
77&$\ket{-{1\over 2},-{1\over 2},{3\over 4},{3\over 4},i,-i,0,0}$&$ {U\over 2} + {{{\sqrt{32\,{t^2} + {U^2} - 4\,{\sqrt{64\,{t^4} + 3\,{t^2}\,{U^2}}}}}}\over 2}$&-3.42151 \\
78&$\ket{-{1\over 2},-{1\over 2},{3\over 4},{3\over 4},1,-i\,{\sqrt{3}},0,0}$&$ {U\over 2} - {{{\sqrt{16\,{t^2} + 4\,t\,U + {U^2}}}}\over 2}$&-8.89512 \\
79&$\ket{-{1\over 2},-{1\over 2},{3\over 4},{3\over 4},1,-i\,{\sqrt{3}},0,0}$&$ {U\over 2} + {{{\sqrt{16\,{t^2} + 4\,t\,U + {U^2}}}}\over 2}$&-1.08488 \\
80&$\ket{-{1\over 2},-{1\over 2},{3\over 4},{3\over 4},1,i\,{\sqrt{3}},0,0}$&$ {U\over 2} - {{{\sqrt{16\,{t^2} + 4\,t\,U + {U^2}}}}\over 2}$&-8.89512 \\
81&$\ket{-{1\over 2},-{1\over 2},{3\over 4},{3\over 4},1,i\,{\sqrt{3}},0,0}$&$ {U\over 2} + {{{\sqrt{16\,{t^2} + 4\,t\,U + {U^2}}}}\over 2}$&-1.08488 \\
\hline \multicolumn{4}{c}{\bf ${\rm N_e}$=3 m$_s$=$-{1\over 2}$ r(r+1)=${3\over 4}$ s(s+1)=${{15}\over 4}$}\\ \hline 
82&$\ket{-{1\over 2},-{1\over 2},{3\over 4},{{15}\over 4},-1,0,0,4}$&$ -2\,t$&-9.49 \\
83&$\ket{-{1\over 2},-{1\over 2},{3\over 4},{{15}\over 4},-i,-2\,i,0,0}$&$ 0$&-7.49 \\
84&$\ket{-{1\over 2},-{1\over 2},{3\over 4},{{15}\over 4},i,2\,i,0,0}$&$ 0$&-7.49 \\
85&$\ket{-{1\over 2},-{1\over 2},{3\over 4},{{15}\over 4},1,0,-4,0}$&$ 2\,t$&-5.49 \\
\hline \multicolumn{4}{c}{\bf ${\rm N_e}$=3 m$_s$=$-{1\over 2}$ r(r+1)=${{15}\over 4}$ s(s+1)=${3\over 4}$}\\ \hline 
86&$\ket{-{1\over 2},-{1\over 2},{{15}\over 4},{3\over 4},-1,0,0,-4}$&$ 2\,t + U$&-0.49 \\
87&$\ket{-{1\over 2},-{1\over 2},{{15}\over 4},{3\over 4},-i,-2\,i,0,0}$&$ U$&-2.49 \\
88&$\ket{-{1\over 2},-{1\over 2},{{15}\over 4},{3\over 4},i,2\,i,0,0}$&$ U$&-2.49 \\
89&$\ket{-{1\over 2},-{1\over 2},{{15}\over 4},{3\over 4},1,0,4,0}$&$ -2\,t + U$&-4.49 \\
\hline \multicolumn{4}{c}{\bf ${\rm N_e}$=3 m$_s$=$-{3\over 2}$ r(r+1)=${3\over 4}$ s(s+1)=${{15}\over 4}$}\\ \hline 
90&$\ket{-{1\over 2},-{3\over 2},{3\over 4},{{15}\over 4},-1,0,0,4}$&$ -2\,t$&-9.47 \\
91&$\ket{-{1\over 2},-{3\over 2},{3\over 4},{{15}\over 4},-i,-2\,i,0,0}$&$ 0$&-7.47 \\
92&$\ket{-{1\over 2},-{3\over 2},{3\over 4},{{15}\over 4},i,2\,i,0,0}$&$ 0$&-7.47 \\
93&$\ket{-{1\over 2},-{3\over 2},{3\over 4},{{15}\over 4},1,0,-4,0}$&$ 2\,t$&-5.47 \\
\end{tabular}
\end{table}
\begin{table}
\caption{Eigenkets and eigenvalues for ${\rm N_e}$=4, spin-up states. 
The abbreviation $\alpha$ is the same as in Tab. \ref{table:n2}}
\begin{tabular}{rlcc}
Number& Eigenstate&\multicolumn{2}{c}{Energy eigenvalue  \hspace{1cm} Value for U/t=5, $\mu$=U/2, and h=0.01t}\\\hline
\hline \multicolumn{4}{c}{\bf ${\rm N_e}$=4 m$_s$=$2$ r(r+1)=$0$ s(s+1)=$6$}\\ \hline 
94&$\ket{0,2,0,6,-1,0,0,4}$&$ 0$&-10.04 \\
\hline \multicolumn{4}{c}{\bf ${\rm N_e}$=4 m$_s$=$1$ r(r+1)=$0$ s(s+1)=$2$}\\ \hline 
95&$\ket{0,1,0,2,-1,-2\,i\,{\sqrt{2}},0,0}$&$ U$&-5.02 \\
96&$\ket{0,1,0,2,-1,2\,i\,{\sqrt{2}},0,0}$&$ U$&-5.02 \\
97&$\ket{0,1,0,2,-i,0,0,0}$&$ {U\over 2} - {{{\sqrt{16\,{t^2} + {U^2}}}}\over 2}$&-10.7216 \\
98&$\ket{0,1,0,2,-i,0,0,0}$&$ {U\over 2} + {{{\sqrt{16\,{t^2} + {U^2}}}}\over 2}$&-4.31844 \\
99&$\ket{0,1,0,2,i,0,0,0}$&$ {U\over 2} - {{{\sqrt{16\,{t^2} + {U^2}}}}\over 2}$&-10.7216 \\
100&$\ket{0,1,0,2,i,0,0,0}$&$ {U\over 2} + {{{\sqrt{16\,{t^2} + {U^2}}}}\over 2}$&-4.31844 \\
101&$\ket{0,1,0,2,1,0,-4,0}$&$ {{2\,U}\over 3} + {{2\,{\sqrt{48\,{t^2} +
     {U^2}}}\,\cos ({{\alpha}\over 3})}\over 3}$&-1.67211 \\
102&$\ket{0,1,0,2,1,0,-4,0}$&$ {{2\,U}\over 3} - {{2\,{\sqrt{48\,{t^2} +
     {U^2}}}\,\cos ({{\pi  - \alpha}\over 3})}\over 3}$&-11.5336 \\
103&$\ket{0,1,0,2,1,0,-4,0}$&$ {{2\,U}\over 3} - {{2\,{\sqrt{48\,{t^2} +
     {U^2}}}\,\cos ({{\pi  + \alpha)}\over 3})}\over 3}$&-6.85429 \\
\hline \multicolumn{4}{c}{\bf ${\rm N_e}$=4 m$_s$=$1$ r(r+1)=$0$ s(s+1)=$6$}\\ \hline 
104&$\ket{0,1,0,6,-1,0,0,4}$&$ 0$&-10.02 \\
\hline \multicolumn{4}{c}{\bf ${\rm N_e}$=4 m$_s$=$1$ r(r+1)=$2$ s(s+1)=$2$}\\ \hline 
105&$\ket{0,1,2,2,-1,0,0,4}$&$ U$&-5.02 \\
106&$\ket{0,1,2,2,-i,-2\,i,0,0}$&$ 2\,t + U$&-3.02 \\
107&$\ket{0,1,2,2,-i,2\,i,0,0}$&$ -2\,t + U$&-7.02 \\
108&$\ket{0,1,2,2,i,-2\,i,0,0}$&$ -2\,t + U$&-7.02 \\
109&$\ket{0,1,2,2,i,2\,i,0,0}$&$ 2\,t + U$&-3.02 \\
110&$\ket{0,1,2,2,1,0,4,0}$&$ U$&-5.02 \\
\end{tabular}
\end{table}
\begin{table}
\caption{Eigenkets and eigenvalues for ${\rm N_e}$=4, states with
spin-projection 0. The abbreviation $\alpha$ is the same as in Tab. \ref{table:n2}, and 
$\beta$ is defined as
$\beta= \arccos({{4\,{t^2}\,U}/{{{\left( {{16\,{t^2}}\over 3} + {{{U^2}}\over
	3} \right) }^{{3\over 2}}}}})$.}
\begin{tabular}{rlcc}
Number& Eigenstate&\multicolumn{2}{c}{Energy eigenvalue  \hspace{1cm} Value for U/t=5, $\mu$=U/2, and h=0.01t}\\\hline
\hline \multicolumn{4}{c}{\bf ${\rm N_e}$=4 m$_s$=$0$ r(r+1)=$0$ s(s+1)=$0$}\\ \hline 
111&$\ket{0,0,0,0,-1,0,0,4}$&$ U - {{2\,{\sqrt{16\,{t^2} + {U^2}}}\,\cos
	({{\beta}\over 3})}\over {{\sqrt{3}}}}$&-11.8443 \\
112&$\ket{0,0,0,0,-1,0,0,4}$&$ U + {{2\,{\sqrt{16\,{t^2} + {U^2}}}\,\cos ({{\pi
	- \beta}\over 3})}\over {{\sqrt{3}}}}$&0.844289 \\
113&$\ket{0,0,0,0,-1,0,0,4}$&$ U + {{2\,{\sqrt{16\,{t^2} + {U^2}}}\,\cos ({{\pi
	+ \beta}\over 3})}\over {{\sqrt{3}}}}$&-4. \\
114&$\ket{0,0,0,0,-i,-2\,i,0,0}$&$ -2\,t + U$&-7. \\
115&$\ket{0,0,0,0,-i,2\,i,0,0}$&$ 2\,t + U$&-3. \\
116&$\ket{0,0,0,0,i,-2\,i,0,0}$&$ 2\,t + U$&-3. \\
117&$\ket{0,0,0,0,i,2\,i,0,0}$&$ -2\,t + U$&-7. \\
118&$\ket{0,0,0,0,1,0,4,0}$&$ U + {{2\,{\sqrt{16\,{t^2} + {U^2}}}\,\cos
	({{\beta}\over 3})}\over {{\sqrt{3}}}}$&1.84429 \\
119&$\ket{0,0,0,0,1,0,4,0}$&$ U - {{2\,{\sqrt{16\,{t^2} + {U^2}}}\,\cos ({{\pi
	- \beta}\over 3})}\over {{\sqrt{3}}}}$&-10.8443 \\
120&$\ket{0,0,0,0,1,0,4,0}$&$ U - {{2\,{\sqrt{16\,{t^2} + {U^2}}}\,\cos ({{\pi
	+ \beta}\over 3})}\over {{\sqrt{3}}}}$&-6. \\
   &&$\beta=\arccos{{4\,{t^2}\,U}\over {{{\left( {{16\,{t^2}}\over 3} + {{{U^2}}\over
	3} \right) }^{{3\over 2}}}}}$&\\
\hline \multicolumn{4}{c}{\bf ${\rm N_e}$=4 m$_s$=$0$ r(r+1)=$0$ s(s+1)=$2$}\\ \hline 
121&$\ket{0,0,0,2,-1,-2\,i\,{\sqrt{2}},0,0}$&$ U$&-5. \\
122&$\ket{0,0,0,2,-1,2\,i\,{\sqrt{2}},0,0}$&$ U$&-5. \\
123&$\ket{0,0,0,2,-i,0,0,0}$&$ {U\over 2} - {{{\sqrt{16\,{t^2} + {U^2}}}}\over 2}$&-10.7016 \\
124&$\ket{0,0,0,2,-i,0,0,0}$&$ {U\over 2} + {{{\sqrt{16\,{t^2} + {U^2}}}}\over 2}$&-4.29844 \\
125&$\ket{0,0,0,2,i,0,0,0}$&$ {U\over 2} - {{{\sqrt{16\,{t^2} + {U^2}}}}\over 2}$&-10.7016 \\
126&$\ket{0,0,0,2,i,0,0,0}$&$ {U\over 2} + {{{\sqrt{16\,{t^2} + {U^2}}}}\over 2}$&-4.29844 \\
127&$\ket{0,0,0,2,1,0,-4,0}$&$ {{2\,U}\over 3} + {{2\,{\sqrt{48\,{t^2} +
	{U^2}}}\,\cos ({{\alpha}\over 3})}\over 3}$&-1.65211 \\
128&$\ket{0,0,0,2,1,0,-4,0}$&$ {{2\,U}\over 3} - {{2\,{\sqrt{48\,{t^2} +
	{U^2}}}\,\cos ({{\pi  - \alpha}\over 3})}\over 3}$&-11.5136 \\
129&$\ket{0,0,0,2,1,0,-4,0}$&$ {{2\,U}\over 3} - {{2\,{\sqrt{48\,{t^2} +
	{U^2}}}\,\cos ({{\pi  + \alpha}\over 3})}\over 3}$&-6.83429 \\
\hline \multicolumn{4}{c}{\bf ${\rm N_e}$=4 m$_s$=$0$ r(r+1)=$0$ s(s+1)=$6$}\\ \hline 
130&$\ket{0,0,0,6,-1,0,0,4}$&$ 0$&-10. \\
\hline \multicolumn{4}{c}{\bf ${\rm N_e}$=4 m$_s$=$0$ r(r+1)=$2$ s(s+1)=$0$}\\ \hline 
131&$\ket{0,0,2,0,-1,0,0,-4}$&$ {{4\,U}\over 3} - {{2\,{\sqrt{48\,{t^2} +
	{U^2}}}\,\cos ({{\alpha)}\over 3})}\over 3}$&-8.34789 \\
132&$\ket{0,0,2,0,-1,0,0,-4}$&$ {{4\,U}\over 3} + {{2\,{\sqrt{48\,{t^2} +
	{U^2}}}\,\cos ({{\pi  - \alpha)}\over 3})}\over 3}$&1.5136 \\
133&$\ket{0,0,2,0,-1,0,0,-4}$&$ {{4\,U}\over 3} + {{2\,{\sqrt{48\,{t^2} +
{U^2}}}\,\cos ({{\pi  + \alpha)}\over 3})}\over 3}$&-3.16571 \\
134&$\ket{0,0,2,0,-i,0,0,0}$&$ {{3\,U}\over 2} - {{{\sqrt{16\,{t^2} + {U^2}}}}\over 2}$&-5.70156 \\
135&$\ket{0,0,2,0,-i,0,0,0}$&$ {{3\,U}\over 2} + {{{\sqrt{16\,{t^2} + {U^2}}}}\over 2}$&0.701562 \\
136&$\ket{0,0,2,0,i,0,0,0}$&$ {{3\,U}\over 2} - {{{\sqrt{16\,{t^2} + {U^2}}}}\over 2}$&-5.70156 \\
137&$\ket{0,0,2,0,i,0,0,0}$&$ {{3\,U}\over 2} + {{{\sqrt{16\,{t^2} + {U^2}}}}\over 2}$&0.701562 \\
138&$\ket{0,0,2,0,1,-2\,i\,{\sqrt{2}},0,0}$&$ U$&-5. \\
139&$\ket{0,0,2,0,1,2\,i\,{\sqrt{2}},0,0}$&$ U$&-5. \\
\hline \multicolumn{4}{c}{\bf ${\rm N_e}$=4 m$_s$=$0$ r(r+1)=$2$ s(s+1)=$2$}\\ \hline 
140&$\ket{0,0,2,2,-1,0,0,4}$&$ U$&-5. \\
141&$\ket{0,0,2,2,-i,-2\,i,0,0}$&$ 2\,t + U$&-3. \\
142&$\ket{0,0,2,2,-i,2\,i,0,0}$&$ -2\,t + U$&-7. \\
143&$\ket{0,0,2,2,i,-2\,i,0,0}$&$ -2\,t + U$&-7. \\
144&$\ket{0,0,2,2,i,2\,i,0,0}$&$ 2\,t + U$&-3. \\
145&$\ket{0,0,2,2,1,0,4,0}$&$ U$&-5. \\
\hline \multicolumn{4}{c}{\bf ${\rm N_e}$=4 m$_s$=$0$ r(r+1)=$6$ s(s+1)=$0$}\\ \hline 
146&$\ket{0,0,6,0,1,0,4,0}$&$ 2\,U$&0 \\
\end{tabular}
\end{table}
\begin{table}
\caption{Eigenkets and eigenvalues for ${\rm N_e}$=4, spin-down
states. The abbreviation $\alpha$ is the same as in Tab. \ref{table:n2}.}
\begin{tabular}{rlcc}
Number& Eigenstate&\multicolumn{2}{c}{Energy eigenvalue  \hspace{1cm} Value for U/t=5, $\mu$=U/2, and h=0.01t}\\\hline
\hline \multicolumn{4}{c}{\bf ${\rm N_e}$=4 m$_s$=$-1$ r(r+1)=$0$ s(s+1)=$2$}\\ \hline 
147&$\ket{0,-1,0,2,-1,-2\,i\,{\sqrt{2}},0,0}$&$ U$&-4.98 \\
148&$\ket{0,-1,0,2,-1,2\,i\,{\sqrt{2}},0,0}$&$ U$&-4.98 \\
149&$\ket{0,-1,0,2,-i,0,0,0}$&$ {U\over 2} - {{{\sqrt{16\,{t^2} + {U^2}}}}\over 2}$&-10.6816 \\
150&$\ket{0,-1,0,2,-i,0,0,0}$&$ {U\over 2} + {{{\sqrt{16\,{t^2} + {U^2}}}}\over 2}$&-4.27844 \\
151&$\ket{0,-1,0,2,i,0,0,0}$&$ {U\over 2} - {{{\sqrt{16\,{t^2} + {U^2}}}}\over 2}$&-10.6816 \\
152&$\ket{0,-1,0,2,i,0,0,0}$&$ {U\over 2} + {{{\sqrt{16\,{t^2} + {U^2}}}}\over 2}$&-4.27844 \\
153&$\ket{0,-1,0,2,1,0,-4,0}$&$ {{2\,U}\over 3} + {{2\,{\sqrt{48\,{t^2} +
	{U^2}}}\,\cos ({{\alpha}\over 3})}\over 3}$&-1.63211 \\
154&$\ket{0,-1,0,2,1,0,-4,0}$&$ {{2\,U}\over 3} - {{2\,{\sqrt{48\,{t^2} +
	{U^2}}}\,\cos ({{\pi  - \alpha}\over 3})}\over 3}$&-11.4936 \\
155&$\ket{0,-1,0,2,1,0,-4,0}$&$ {{2\,U}\over 3} - {{2\,{\sqrt{48\,{t^2} +
	{U^2}}}\,\cos ({{\pi  + \alpha}\over 3})}\over 3}$&-6.81429 \\
\hline \multicolumn{4}{c}{\bf ${\rm N_e}$=4 m$_s$=$-1$ r(r+1)=$0$ s(s+1)=$6$}\\ \hline 
156&$\ket{0,-1,0,6,-1,0,0,4}$&$ 0$&-9.98 \\
\hline \multicolumn{4}{c}{\bf ${\rm N_e}$=4 m$_s$=$-1$ r(r+1)=$2$ s(s+1)=$2$}\\ \hline 
157&$\ket{0,-1,2,2,-1,0,0,4}$&$ U$&-4.98 \\
158&$\ket{0,-1,2,2,-i,-2\,i,0,0}$&$ 2\,t + U$&-2.98 \\
159&$\ket{0,-1,2,2,-i,2\,i,0,0}$&$ -2\,t + U$&-6.98 \\
160&$\ket{0,-1,2,2,i,-2\,i,0,0}$&$ -2\,t + U$&-6.98 \\
161&$\ket{0,-1,2,2,i,2\,i,0,0}$&$ 2\,t + U$&-2.98 \\
162&$\ket{0,-1,2,2,1,0,4,0}$&$ U$&-4.98 \\
\hline \multicolumn{4}{c}{\bf ${\rm N_e}$=4 m$_s$=$-2$ r(r+1)=$0$ s(s+1)=$6$}\\ \hline 
163&$\ket{0,-2,0,6,-1,0,0,4}$&$ 0$&-9.96 \\
\end{tabular}
\end{table}
\begin{table}\caption{Eigenkets and eigenvalues for ${\rm N_e}$=5, spin-up states.}
\begin{tabular}{rlcc}
Number& Eigenstate&\multicolumn{2}{c}{Energy eigenvalue  \hspace{1cm} Value for U/t=5, $\mu$=U/2, and h=0.01t}\\\hline
\hline \multicolumn{4}{c}{\bf ${\rm N_e}$=5 m$_s$=${3\over 2}$ r(r+1)=${3\over 4}$ s(s+1)=${{15}\over 4}$}\\ \hline 
164&$\ket{{1\over 2},{3\over 2},{3\over 4},{{15}\over 4},-1,0,0,4}$&$ 2\,t + U$&-5.53 \\
165&$\ket{{1\over 2},{3\over 2},{3\over 4},{{15}\over 4},-i,2\,i,0,0}$&$ U$&-7.53 \\
166&$\ket{{1\over 2},{3\over 2},{3\over 4},{{15}\over 4},i,-2\,i,0,0}$&$ U$&-7.53 \\
167&$\ket{{1\over 2},{3\over 2},{3\over 4},{{15}\over 4},1,0,-4,0}$&$ -2\,t + U$&-9.53 \\
\hline \multicolumn{4}{c}{\bf ${\rm N_e}$=5 m$_s$=${1\over 2}$ r(r+1)=${3\over 4}$ s(s+1)=${3\over 4}$}\\ \hline 
168&$\ket{{1\over 2},{1\over 2},{3\over 4},{3\over 4},-1,-i\,{\sqrt{3}},0,0}$&$ {{3\,U}\over 2} - {{{\sqrt{16\,{t^2} + 4\,t\,U + {U^2}}}}\over 2}$&-8.91512 \\
169&$\ket{{1\over 2},{1\over 2},{3\over 4},{3\over 4},-1,-i\,{\sqrt{3}},0,0}$&$ {{3\,U}\over 2} + {{{\sqrt{16\,{t^2} + 4\,t\,U + {U^2}}}}\over 2}$&-1.10488 \\
170&$\ket{{1\over 2},{1\over 2},{3\over 4},{3\over 4},-1,i\,{\sqrt{3}},0,0}$&$ {{3\,U}\over 2} - {{{\sqrt{16\,{t^2} + 4\,t\,U + {U^2}}}}\over 2}$&-8.91512 \\
171&$\ket{{1\over 2},{1\over 2},{3\over 4},{3\over 4},-1,i\,{\sqrt{3}},0,0}$&$ {{3\,U}\over 2} + {{{\sqrt{16\,{t^2} + 4\,t\,U + {U^2}}}}\over 2}$&-1.10488 \\
172&$\ket{{1\over 2},{1\over 2},{3\over 4},{3\over 4},-i,-i,0,0}$&$ {{3\,U}\over 2} - {{{\sqrt{32\,{t^2} + {U^2} + 4\,{\sqrt{64\,{t^4} + 3\,{t^2}\,{U^2}}}}}}\over 2}$&-10.1129 \\
173&$\ket{{1\over 2},{1\over 2},{3\over 4},{3\over 4},-i,-i,0,0}$&$ {{3\,U}\over 2} + {{{\sqrt{32\,{t^2} + {U^2} + 4\,{\sqrt{64\,{t^4} + 3\,{t^2}\,{U^2}}}}}}\over 2}$&0.0929233 \\
174&$\ket{{1\over 2},{1\over 2},{3\over 4},{3\over 4},-i,-i,0,0}$&$ {{3\,U}\over 2} - {{{\sqrt{32\,{t^2} + {U^2} - 4\,{\sqrt{64\,{t^4} + 3\,{t^2}\,{U^2}}}}}}\over 2}$&-6.57849 \\
175&$\ket{{1\over 2},{1\over 2},{3\over 4},{3\over 4},-i,-i,0,0}$&$ {{3\,U}\over 2} + {{{\sqrt{32\,{t^2} + {U^2} - 4\,{\sqrt{64\,{t^4} + 3\,{t^2}\,{U^2}}}}}}\over 2}$&-3.44151 \\
176&$\ket{{1\over 2},{1\over 2},{3\over 4},{3\over 4},i,i,0,0}$&$ {{3\,U}\over 2} - {{{\sqrt{32\,{t^2} + {U^2} + 4\,{\sqrt{64\,{t^4} + 3\,{t^2}\,{U^2}}}}}}\over 2}$&-10.1129 \\
177&$\ket{{1\over 2},{1\over 2},{3\over 4},{3\over 4},i,i,0,0}$&$ {{3\,U}\over 2} + {{{\sqrt{32\,{t^2} + {U^2} + 4\,{\sqrt{64\,{t^4} + 3\,{t^2}\,{U^2}}}}}}\over 2}$&0.0929233 \\
178&$\ket{{1\over 2},{1\over 2},{3\over 4},{3\over 4},i,i,0,0}$&$ {{3\,U}\over 2} - {{{\sqrt{32\,{t^2} + {U^2} - 4\,{\sqrt{64\,{t^4} + 3\,{t^2}\,{U^2}}}}}}\over 2}$&-6.57849 \\
179&$\ket{{1\over 2},{1\over 2},{3\over 4},{3\over 4},i,i,0,0}$&$ {{3\,U}\over 2} + {{{\sqrt{32\,{t^2} + {U^2} - 4\,{\sqrt{64\,{t^4} + 3\,{t^2}\,{U^2}}}}}}\over 2}$&-3.44151 \\
180&$\ket{{1\over 2},{1\over 2},{3\over 4},{3\over 4},1,-i\,{\sqrt{3}},0,0}$&$ {{3\,U}\over 2} - {{{\sqrt{16\,{t^2} - 4\,t\,U + {U^2}}}}\over 2}$&-7.30129 \\
181&$\ket{{1\over 2},{1\over 2},{3\over 4},{3\over 4},1,-i\,{\sqrt{3}},0,0}$&$ {{3\,U}\over 2} + {{{\sqrt{16\,{t^2} - 4\,t\,U + {U^2}}}}\over 2}$&-2.71871 \\
182&$\ket{{1\over 2},{1\over 2},{3\over 4},{3\over 4},1,i\,{\sqrt{3}},0,0}$&$ {{3\,U}\over 2} - {{{\sqrt{16\,{t^2} - 4\,t\,U + {U^2}}}}\over 2}$&-7.30129 \\
183&$\ket{{1\over 2},{1\over 2},{3\over 4},{3\over 4},1,i\,{\sqrt{3}},0,0}$&$ {{3\,U}\over 2} + {{{\sqrt{16\,{t^2} - 4\,t\,U + {U^2}}}}\over 2}$&-2.71871 \\
\hline \multicolumn{4}{c}{\bf ${\rm N_e}$=5 m$_s$=${1\over 2}$ r(r+1)=${3\over 4}$ s(s+1)=${{15}\over 4}$}\\ \hline 
184&$\ket{{1\over 2},{1\over 2},{3\over 4},{{15}\over 4},-1,0,0,4}$&$ 2\,t + U$&-5.51 \\
185&$\ket{{1\over 2},{1\over 2},{3\over 4},{{15}\over 4},-i,2\,i,0,0}$&$ U$&-7.51 \\
186&$\ket{{1\over 2},{1\over 2},{3\over 4},{{15}\over 4},i,-2\,i,0,0}$&$ U$&-7.51 \\
187&$\ket{{1\over 2},{1\over 2},{3\over 4},{{15}\over 4},1,0,-4,0}$&$ -2\,t + U$&-9.51 \\
\hline \multicolumn{4}{c}{\bf ${\rm N_e}$=5 m$_s$=${1\over 2}$ r(r+1)=${{15}\over 4}$ s(s+1)=${3\over 4}$}\\ \hline 
188&$\ket{{1\over 2},{1\over 2},{{15}\over 4},{3\over 4},-1,0,0,-4}$&$ -2\,t + 2\,U$&-4.51 \\
189&$\ket{{1\over 2},{1\over 2},{{15}\over 4},{3\over 4},-i,2\,i,0,0}$&$ 2\,U$&-2.51 \\
190&$\ket{{1\over 2},{1\over 2},{{15}\over 4},{3\over 4},i,-2\,i,0,0}$&$ 2\,U$&-2.51 \\
191&$\ket{{1\over 2},{1\over 2},{{15}\over 4},{3\over 4},1,0,4,0}$&$ 2\,t + 2\,U$&-0.51 \\
\end{tabular}
\end{table}
\begin{table}\caption{Eigenkets and eigenvalues for ${\rm N_e}$=5, spin-down states.}
\begin{tabular}{rlcc}
Number& Eigenstate&\multicolumn{2}{c}{Energy eigenvalue  \hspace{1cm} Value for U/t=5, $\mu$=U/2, and h=0.01t}\\\hline
\hline \multicolumn{4}{c}{\bf ${\rm N_e}$=5 m$_s$=$-{1\over 2}$ r(r+1)=${3\over 4}$ s(s+1)=${3\over 4}$}\\ \hline 
192&$\ket{{1\over 2},-{1\over 2},{3\over 4},{3\over 4},-1,-i\,{\sqrt{3}},0,0}$&$ {{3\,U}\over 2} - {{{\sqrt{16\,{t^2} + 4\,t\,U + {U^2}}}}\over 2}$&-8.89512 \\
193&$\ket{{1\over 2},-{1\over 2},{3\over 4},{3\over 4},-1,-i\,{\sqrt{3}},0,0}$&$ {{3\,U}\over 2} + {{{\sqrt{16\,{t^2} + 4\,t\,U + {U^2}}}}\over 2}$&-1.08488 \\
194&$\ket{{1\over 2},-{1\over 2},{3\over 4},{3\over 4},-1,i\,{\sqrt{3}},0,0}$&$ {{3\,U}\over 2} - {{{\sqrt{16\,{t^2} + 4\,t\,U + {U^2}}}}\over 2}$&-8.89512 \\
195&$\ket{{1\over 2},-{1\over 2},{3\over 4},{3\over 4},-1,i\,{\sqrt{3}},0,0}$&$ {{3\,U}\over 2} + {{{\sqrt{16\,{t^2} + 4\,t\,U + {U^2}}}}\over 2}$&-1.08488 \\
196&$\ket{{1\over 2},-{1\over 2},{3\over 4},{3\over 4},-i,-i,0,0}$&$ {{3\,U}\over 2} - {{{\sqrt{32\,{t^2} + {U^2} + 4\,{\sqrt{64\,{t^4} + 3\,{t^2}\,{U^2}}}}}}\over 2}$&-10.0929 \\
197&$\ket{{1\over 2},-{1\over 2},{3\over 4},{3\over 4},-i,-i,0,0}$&$ {{3\,U}\over 2} + {{{\sqrt{32\,{t^2} + {U^2} + 4\,{\sqrt{64\,{t^4} + 3\,{t^2}\,{U^2}}}}}}\over 2}$&0.112923 \\
198&$\ket{{1\over 2},-{1\over 2},{3\over 4},{3\over 4},-i,-i,0,0}$&$ {{3\,U}\over 2} - {{{\sqrt{32\,{t^2} + {U^2} - 4\,{\sqrt{64\,{t^4} + 3\,{t^2}\,{U^2}}}}}}\over 2}$&-6.55849 \\
199&$\ket{{1\over 2},-{1\over 2},{3\over 4},{3\over 4},-i,-i,0,0}$&$ {{3\,U}\over 2} + {{{\sqrt{32\,{t^2} + {U^2} - 4\,{\sqrt{64\,{t^4} + 3\,{t^2}\,{U^2}}}}}}\over 2}$&-3.42151 \\
200&$\ket{{1\over 2},-{1\over 2},{3\over 4},{3\over 4},i,i,0,0}$&$ {{3\,U}\over 2} - {{{\sqrt{32\,{t^2} + {U^2} + 4\,{\sqrt{64\,{t^4} + 3\,{t^2}\,{U^2}}}}}}\over 2}$&-10.0929 \\
201&$\ket{{1\over 2},-{1\over 2},{3\over 4},{3\over 4},i,i,0,0}$&$ {{3\,U}\over 2} + {{{\sqrt{32\,{t^2} + {U^2} + 4\,{\sqrt{64\,{t^4} + 3\,{t^2}\,{U^2}}}}}}\over 2}$&0.112923 \\
202&$\ket{{1\over 2},-{1\over 2},{3\over 4},{3\over 4},i,i,0,0}$&$ {{3\,U}\over 2} - {{{\sqrt{32\,{t^2} + {U^2} - 4\,{\sqrt{64\,{t^4} + 3\,{t^2}\,{U^2}}}}}}\over 2}$&-6.55849 \\
203&$\ket{{1\over 2},-{1\over 2},{3\over 4},{3\over 4},i,i,0,0}$&$ {{3\,U}\over 2} + {{{\sqrt{32\,{t^2} + {U^2} - 4\,{\sqrt{64\,{t^4} + 3\,{t^2}\,{U^2}}}}}}\over 2}$&-3.42151 \\
204&$\ket{{1\over 2},-{1\over 2},{3\over 4},{3\over 4},1,-i\,{\sqrt{3}},0,0}$&$ {{3\,U}\over 2} - {{{\sqrt{16\,{t^2} - 4\,t\,U + {U^2}}}}\over 2}$&-7.28129 \\
205&$\ket{{1\over 2},-{1\over 2},{3\over 4},{3\over 4},1,-i\,{\sqrt{3}},0,0}$&$ {{3\,U}\over 2} + {{{\sqrt{16\,{t^2} - 4\,t\,U + {U^2}}}}\over 2}$&-2.69871 \\
206&$\ket{{1\over 2},-{1\over 2},{3\over 4},{3\over 4},1,i\,{\sqrt{3}},0,0}$&$ {{3\,U}\over 2} - {{{\sqrt{16\,{t^2} - 4\,t\,U + {U^2}}}}\over 2}$&-7.28129 \\
207&$\ket{{1\over 2},-{1\over 2},{3\over 4},{3\over 4},1,i\,{\sqrt{3}},0,0}$&$ {{3\,U}\over 2} + {{{\sqrt{16\,{t^2} - 4\,t\,U + {U^2}}}}\over 2}$&-2.69871 \\
\hline \multicolumn{4}{c}{\bf ${\rm N_e}$=5 m$_s$=$-{1\over 2}$ r(r+1)=${3\over 4}$ s(s+1)=${{15}\over 4}$}\\ \hline 
208&$\ket{{1\over 2},-{1\over 2},{3\over 4},{{15}\over 4},-1,0,0,4}$&$ 2\,t + U$&-5.49 \\
209&$\ket{{1\over 2},-{1\over 2},{3\over 4},{{15}\over 4},-i,2\,i,0,0}$&$ U$&-7.49 \\
210&$\ket{{1\over 2},-{1\over 2},{3\over 4},{{15}\over 4},i,-2\,i,0,0}$&$ U$&-7.49 \\
211&$\ket{{1\over 2},-{1\over 2},{3\over 4},{{15}\over 4},1,0,-4,0}$&$ -2\,t + U$&-9.49 \\
\hline \multicolumn{4}{c}{\bf ${\rm N_e}$=5 m$_s$=$-{1\over 2}$ r(r+1)=${{15}\over 4}$ s(s+1)=${3\over 4}$}\\ \hline 
212&$\ket{{1\over 2},-{1\over 2},{{15}\over 4},{3\over 4},-1,0,0,-4}$&$ -2\,t + 2\,U$&-4.49 \\
213&$\ket{{1\over 2},-{1\over 2},{{15}\over 4},{3\over 4},-i,2\,i,0,0}$&$ 2\,U$&-2.49 \\
214&$\ket{{1\over 2},-{1\over 2},{{15}\over 4},{3\over 4},i,-2\,i,0,0}$&$ 2\,U$&-2.49 \\
215&$\ket{{1\over 2},-{1\over 2},{{15}\over 4},{3\over 4},1,0,4,0}$&$ 2\,t + 2\,U$&-0.49 \\
\hline \multicolumn{4}{c}{\bf ${\rm N_e}$=5 m$_s$=$-{3\over 2}$ r(r+1)=${3\over 4}$ s(s+1)=${{15}\over 4}$}\\ \hline 
216&$\ket{{1\over 2},-{3\over 2},{3\over 4},{{15}\over 4},-1,0,0,4}$&$ 2\,t + U$&-5.47 \\
217&$\ket{{1\over 2},-{3\over 2},{3\over 4},{{15}\over 4},-i,2\,i,0,0}$&$ U$&-7.47 \\
218&$\ket{{1\over 2},-{3\over 2},{3\over 4},{{15}\over 4},i,-2\,i,0,0}$&$ U$&-7.47 \\
219&$\ket{{1\over 2},-{3\over 2},{3\over 4},{{15}\over 4},1,0,-4,0}$&$ -2\,t + U$&-9.47 \\
\end{tabular}
\end{table}
\begin{table}
\caption{Eigenkets and eigenvalues for ${\rm N_e}$=6.
The abbreviation $\alpha$ is the same as in Tab. \ref{table:n2}.}
\begin{tabular}{rlcc}
Number& Eigenstate&\multicolumn{2}{c}{Energy eigenvalue  \hspace{1cm} Value for U/t=5, $\mu$=U/2, and h=0.01t}\\\hline
\hline \multicolumn{4}{c}{\bf ${\rm N_e}$=6 m$_s$=$1$ r(r+1)=$2$ s(s+1)=$2$}\\ \hline 
220&$\ket{1,1,2,2,-1,0,0,-4}$&$ 2\,U$&-5.02 \\
221&$\ket{1,1,2,2,-i,-2\,i,0,0}$&$ -2\,t + 2\,U$&-7.02 \\
222&$\ket{1,1,2,2,-i,2\,i,0,0}$&$ 2\,t + 2\,U$&-3.02 \\
223&$\ket{1,1,2,2,i,-2\,i,0,0}$&$ 2\,t + 2\,U$&-3.02 \\
224&$\ket{1,1,2,2,i,2\,i,0,0}$&$ -2\,t + 2\,U$&-7.02 \\
225&$\ket{1,1,2,2,1,0,-4,0}$&$ 2\,U$&-5.02 \\
\hline \multicolumn{4}{c}{\bf ${\rm N_e}$=6 m$_s$=$0$ r(r+1)=$2$ s(s+1)=$0$}\\ \hline 
226&$\ket{1,0,2,0,-1,-2\,i\,{\sqrt{2}},0,0}$&$ 2\,U$&-5. \\
227&$\ket{1,0,2,0,-1,2\,i\,{\sqrt{2}},0,0}$&$ 2\,U$&-5. \\
228&$\ket{1,0,2,0,-i,0,0,0}$&$ {{5\,U}\over 2} - {{{\sqrt{16\,{t^2} + {U^2}}}}\over 2}$&-5.70156 \\
229&$\ket{1,0,2,0,-i,0,0,0}$&$ {{5\,U}\over 2} + {{{\sqrt{16\,{t^2} + {U^2}}}}\over 2}$&0.701562 \\
230&$\ket{1,0,2,0,i,0,0,0}$&$ {{5\,U}\over 2} - {{{\sqrt{16\,{t^2} + {U^2}}}}\over 2}$&-5.70156 \\
231&$\ket{1,0,2,0,i,0,0,0}$&$ {{5\,U}\over 2} + {{{\sqrt{16\,{t^2} + {U^2}}}}\over 2}$&0.701562 \\
232&$\ket{1,0,2,0,1,0,4,0}$&$ {{7\,U}\over 3} - {{2\,{\sqrt{48\,{t^2} +
	{U^2}}}\,\cos ({{\alpha}\over 3})}\over 3}$&-8.34789 \\
233&$\ket{1,0,2,0,1,0,4,0}$&$ {{7\,U}\over 3} + {{2\,{\sqrt{48\,{t^2} +
	{U^2}}}\,\cos ({{\pi  - \alpha}\over 3})}\over 3}$&1.5136 \\
234&$\ket{1,0,2,0,1,0,4,0}$&$ {{7\,U}\over 3} + {{2\,{\sqrt{48\,{t^2} +
	{U^2}}}\,\cos ({{\pi  + \alpha}\over 3})}\over 3}$&-3.16571 \\
\hline \multicolumn{4}{c}{\bf ${\rm N_e}$=6 m$_s$=$0$ r(r+1)=$2$ s(s+1)=$2$}\\ \hline 
235&$\ket{1,0,2,2,-1,0,0,-4}$&$ 2\,U$&-5. \\
236&$\ket{1,0,2,2,-i,-2\,i,0,0}$&$ -2\,t + 2\,U$&-7. \\
237&$\ket{1,0,2,2,-i,2\,i,0,0}$&$ 2\,t + 2\,U$&-3. \\
238&$\ket{1,0,2,2,i,-2\,i,0,0}$&$ 2\,t + 2\,U$&-3. \\
239&$\ket{1,0,2,2,i,2\,i,0,0}$&$ -2\,t + 2\,U$&-7. \\
240&$\ket{1,0,2,2,1,0,-4,0}$&$ 2\,U$&-5. \\
\hline \multicolumn{4}{c}{\bf ${\rm N_e}$=6 m$_s$=$0$ r(r+1)=$6$ s(s+1)=$0$}\\ \hline 
241&$\ket{1,0,6,0,-1,0,0,-4}$&$ 3\,U$&0 \\
\hline \multicolumn{4}{c}{\bf ${\rm N_e}$=6 m$_s$=$-1$ r(r+1)=$2$ s(s+1)=$2$}\\ \hline 
242&$\ket{1,-1,2,2,-1,0,0,-4}$&$ 2\,U$&-4.98 \\
243&$\ket{1,-1,2,2,-i,-2\,i,0,0}$&$ -2\,t + 2\,U$&-6.98 \\
244&$\ket{1,-1,2,2,-i,2\,i,0,0}$&$ 2\,t + 2\,U$&-2.98 \\
245&$\ket{1,-1,2,2,i,-2\,i,0,0}$&$ 2\,t + 2\,U$&-2.98 \\
246&$\ket{1,-1,2,2,i,2\,i,0,0}$&$ -2\,t + 2\,U$&-6.98 \\
247&$\ket{1,-1,2,2,1,0,-4,0}$&$ 2\,U$&-4.98 \\
\end{tabular}
\end{table}
\begin{table}\caption{Eigenkets and eigenvalues for ${\rm N_e}$=7.}
\begin{tabular}{rlcc}
Number& Eigenstate&\multicolumn{2}{c}{Energy eigenvalue  \hspace{1cm} Value for U/t=5, $\mu$=U/2, and h=0.01t}\\\hline
\hline \multicolumn{4}{c}{\bf ${\rm N_e}$=7 m$_s$=${1\over 2}$ r(r+1)=${{15}\over 4}$ s(s+1)=${3\over 4}$}\\ \hline 
248&$\ket{{3\over 2},{1\over 2},{{15}\over 4},{3\over 4},-1,0,0,-4}$&$ 2\,t + 3\,U$&-0.51 \\
249&$\ket{{3\over 2},{1\over 2},{{15}\over 4},{3\over 4},-i,-2\,i,0,0}$&$ 3\,U$&-2.51 \\
250&$\ket{{3\over 2},{1\over 2},{{15}\over 4},{3\over 4},i,2\,i,0,0}$&$ 3\,U$&-2.51 \\
251&$\ket{{3\over 2},{1\over 2},{{15}\over 4},{3\over 4},1,0,4,0}$&$ -2\,t + 3\,U$&-4.51 \\
\hline \multicolumn{4}{c}{\bf ${\rm N_e}$=7 m$_s$=$-{1\over 2}$ r(r+1)=${{15}\over 4}$ s(s+1)=${3\over 4}$}\\ \hline 
252&$\ket{{3\over 2},-{1\over 2},{{15}\over 4},{3\over 4},-1,0,0,-4}$&$ 2\,t + 3\,U$&-0.49 \\
253&$\ket{{3\over 2},-{1\over 2},{{15}\over 4},{3\over 4},-i,-2\,i,0,0}$&$ 3\,U$&-2.49 \\
254&$\ket{{3\over 2},-{1\over 2},{{15}\over 4},{3\over 4},i,2\,i,0,0}$&$ 3\,U$&-2.49 \\
255&$\ket{{3\over 2},-{1\over 2},{{15}\over 4},{3\over 4},1,0,4,0}$&$ -2\,t + 3\,U$&-4.49 \\
\end{tabular}
\end{table}
\begin{table}\caption{Eigenkets and eigenvalues for ${\rm N_e}$=8.}
\begin{tabular}{rlcc}
Number& Eigenstate&\multicolumn{2}{c}{Energy eigenvalue  \hspace{1cm} Value for U/t=5, $\mu$=U/2, and h=0.01t}\\\hline
\hline \multicolumn{4}{c}{\bf ${\rm N_e}$=8 m$_s$=$0$ r(r+1)=$6$ s(s+1)=$0$}\\ \hline 
256&$\ket{2,0,6,0,1,0,4,0}$&$ 4\,U$&0 \\
\end{tabular}
\label{table:n8}
\end{table}
\end{multicols}
\clearpage
\begin{center}
\thispagestyle{empty}
\begin{minipage}[t]{12cm}
\psfig{file=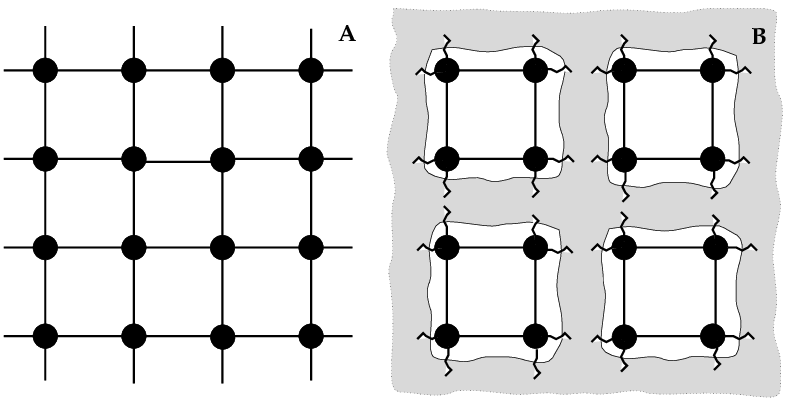,width=8.5cm}
\end{minipage}\\ \vspace*{\fill} Figure 1
\newpage
\thispagestyle{empty}
\begin{minipage}[t]{12cm}
\psfig{file=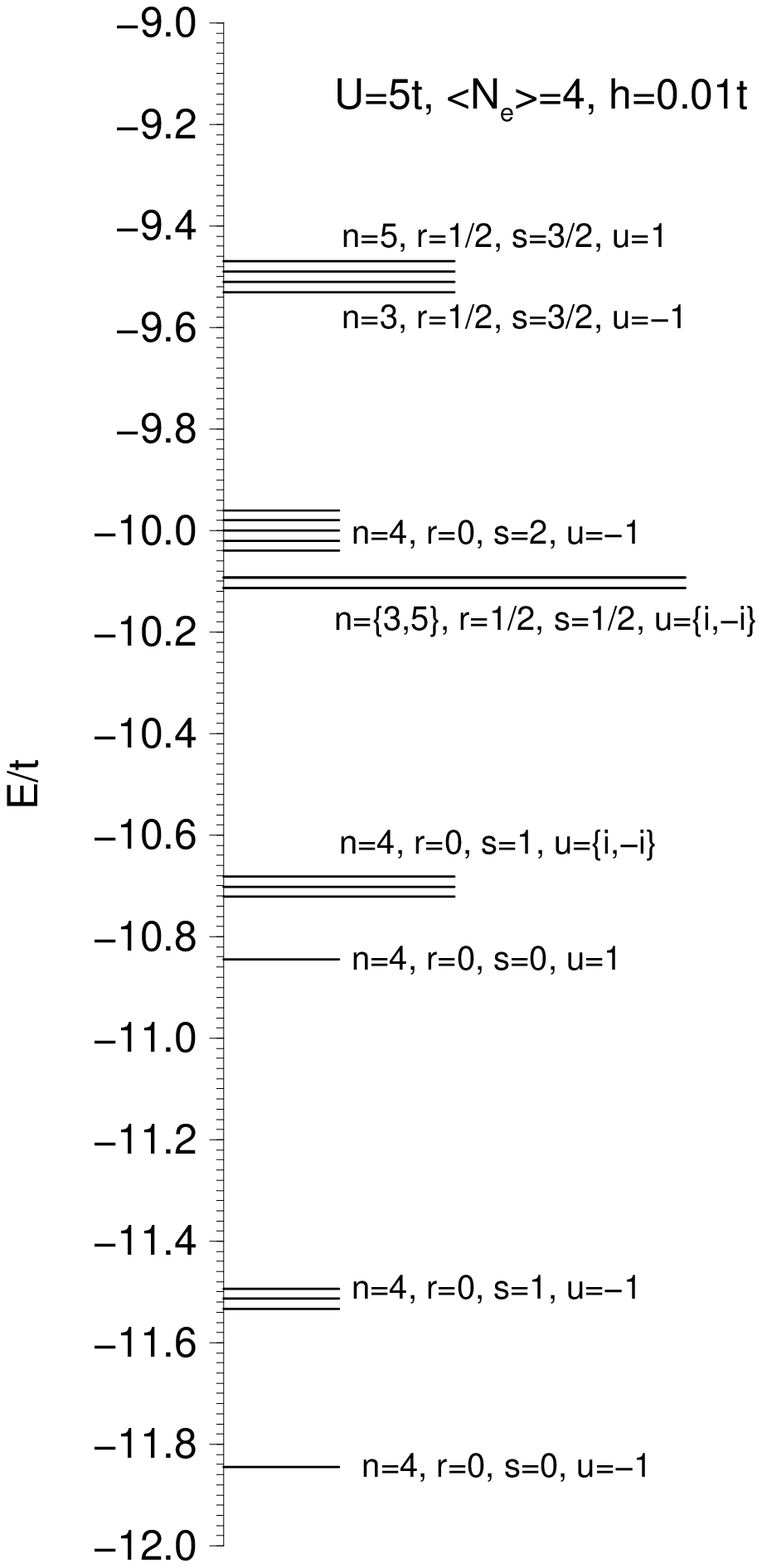,width=8cm}
\end{minipage}\\ \vspace*{\fill} Figure 2
\newpage
\thispagestyle{empty}
\begin{minipage}[t]{12cm}
\psfig{file=./figure03.eps,width=8cm}
\end{minipage}\\ \vspace*{\fill} Figure 3
\newpage
\thispagestyle{empty}
\begin{minipage}[t]{12cm}
\psfig{file=./figure04.eps,width=8cm}
\end{minipage}\\ \vspace*{\fill} Figure 4
\newpage
\thispagestyle{empty}
\begin{minipage}[t]{12cm}
\psfig{file=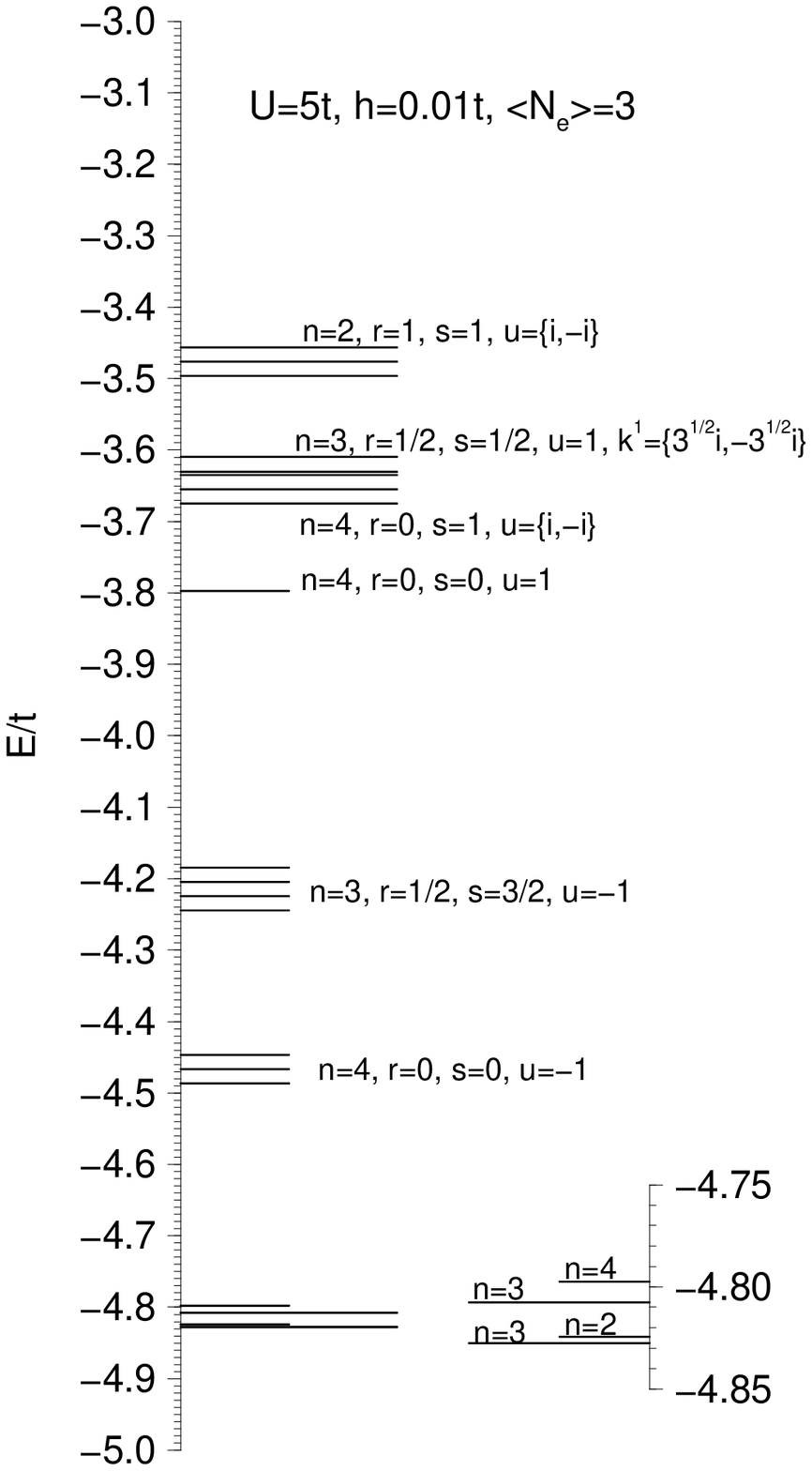,width=8cm}
\end{minipage}\\ \vspace*{\fill} Figure 5
\newpage
\thispagestyle{empty}
\begin{minipage}[t]{12cm}
\psfig{file=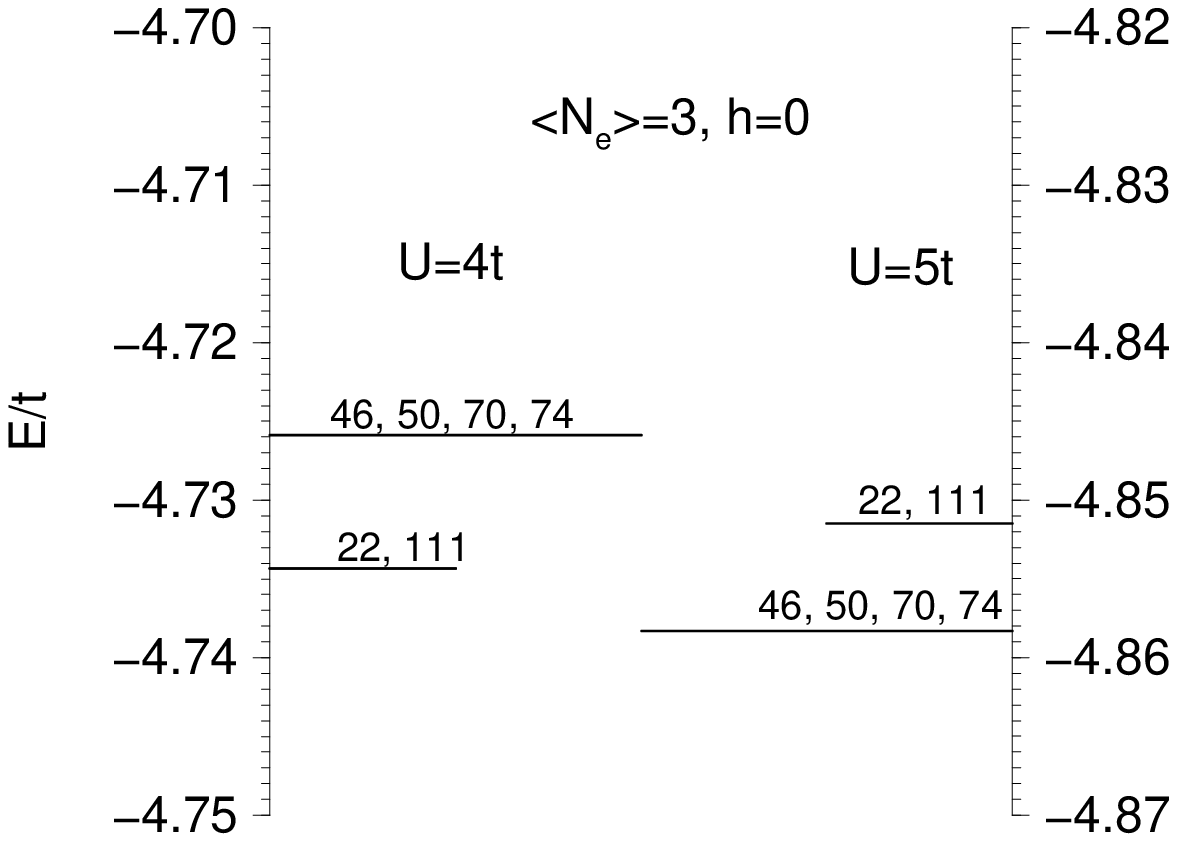,width=8cm}
\end{minipage}\\ \vspace*{\fill} Figure 6
\newpage
\thispagestyle{empty}
\begin{minipage}[t]{12cm}
\psfig{file=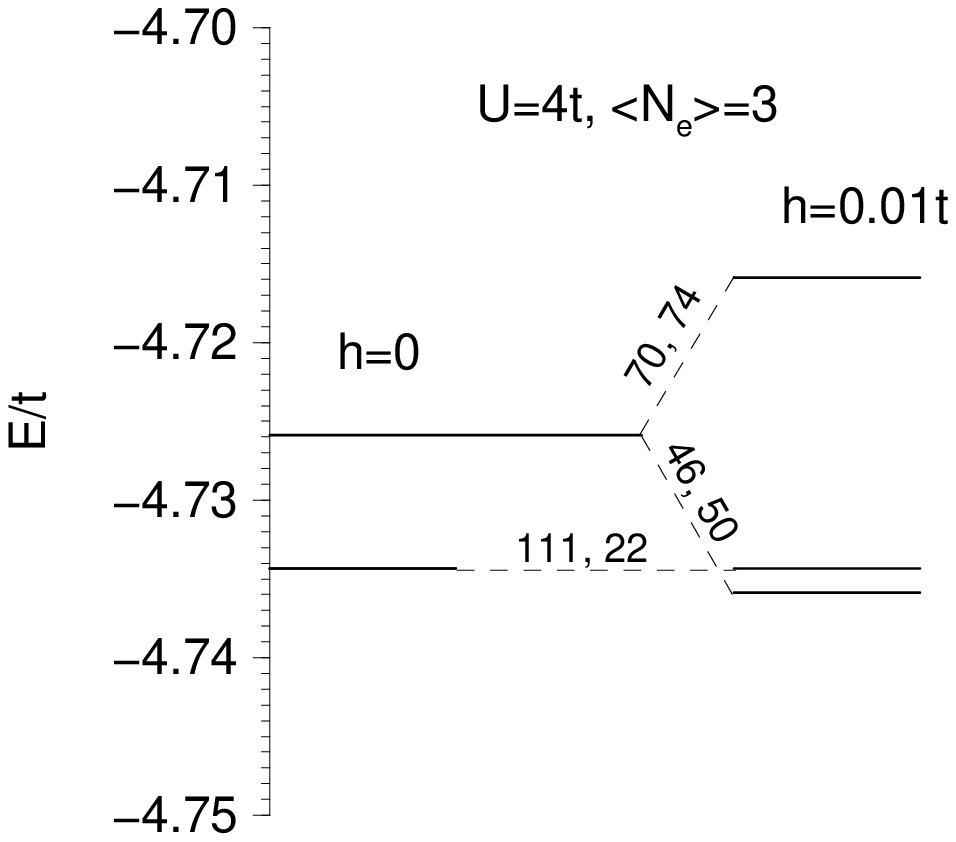,width=8cm}
\end{minipage}\\ \vspace*{\fill} Figure 7
\newpage
\thispagestyle{empty}
\begin{minipage}[t]{12cm}
\psfig{file=./figure08.eps,width=8cm}
\end{minipage}\\ \vspace*{\fill} Figure 8
\newpage
\thispagestyle{empty}
\begin{minipage}[t]{12cm}
\psfig{file=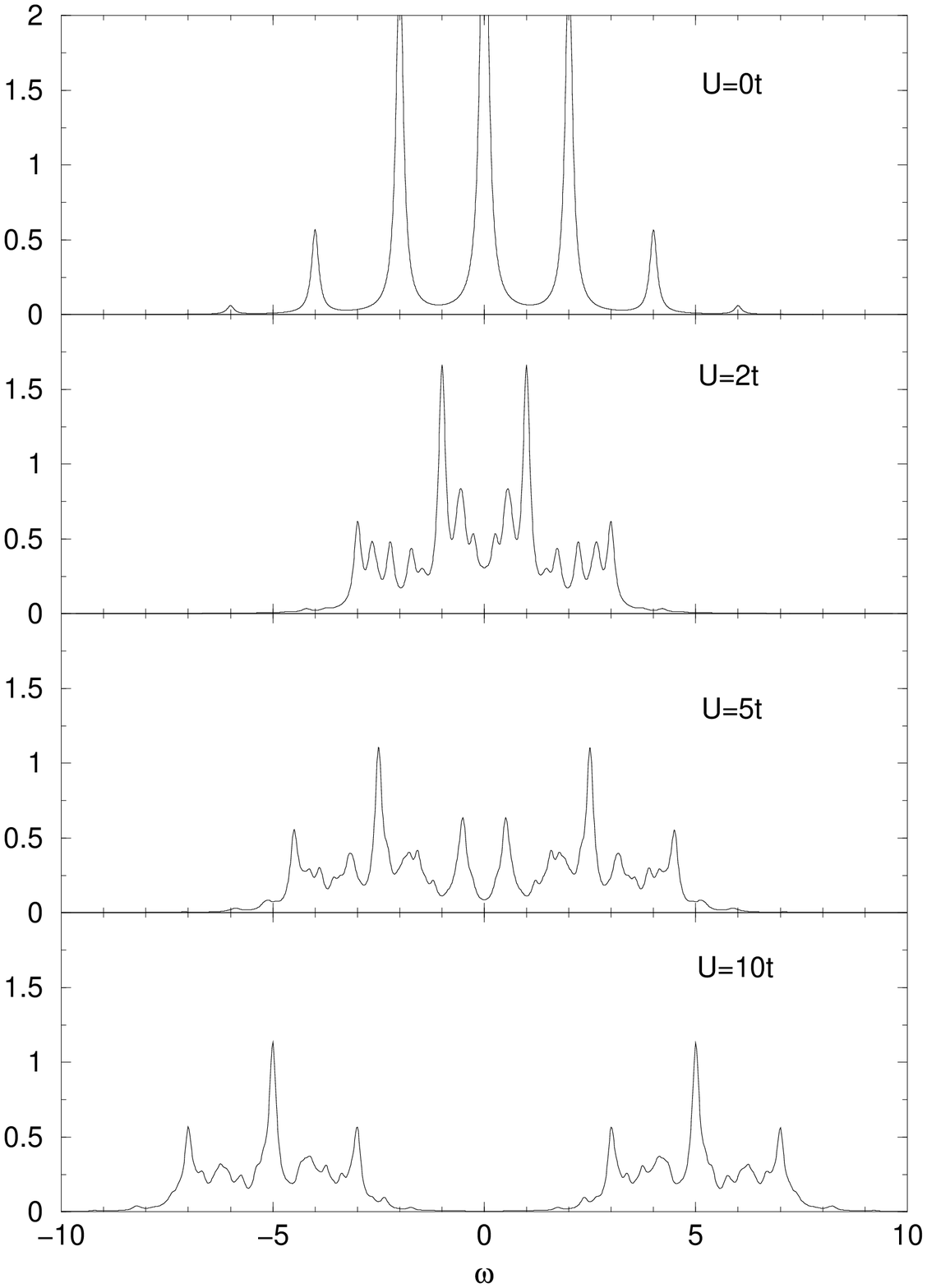,width=8cm}
\end{minipage}\\ \vspace*{\fill} Figure 9
\newpage
\thispagestyle{empty}
\begin{minipage}[t]{12cm}
\psfig{file=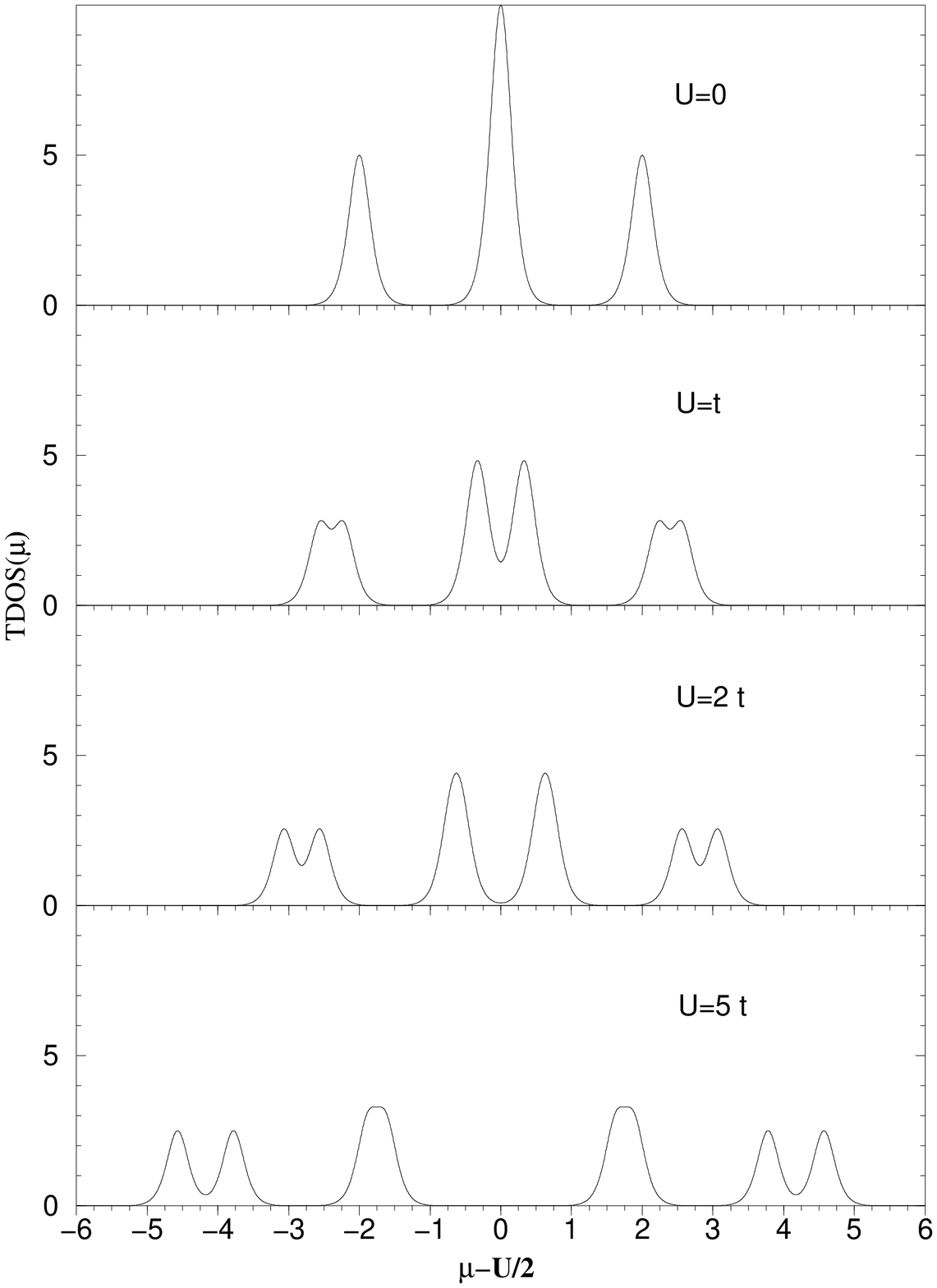,width=8cm}
\end{minipage}\\ \vspace*{\fill} Figure 10
\newpage
\thispagestyle{empty}
\begin{minipage}[t]{12cm}
\psfig{file=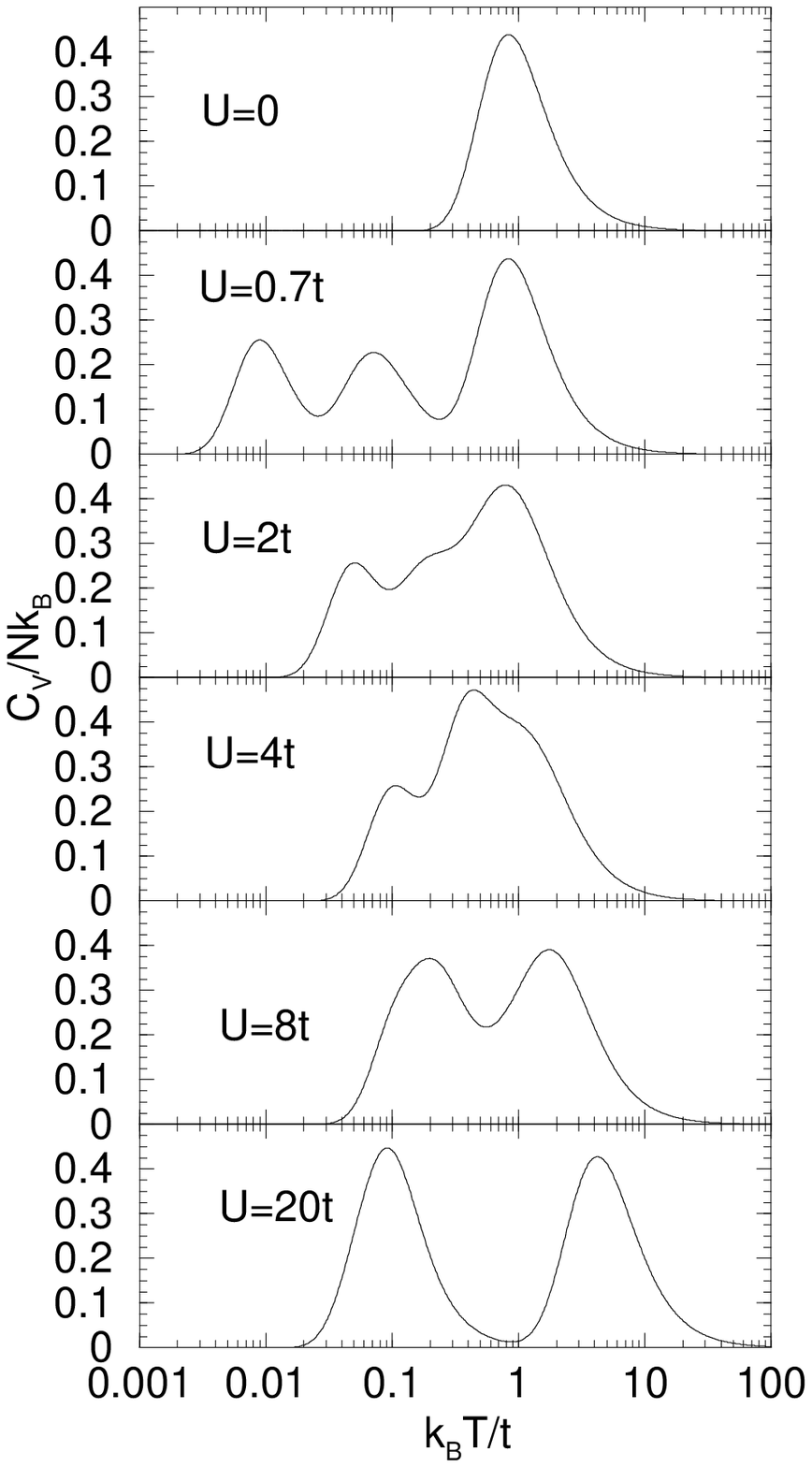,width=8cm}
\end{minipage}\\ \vspace*{\fill} Figure 11
\newpage
\thispagestyle{empty}
\begin{minipage}[t]{12cm}
\psfig{file=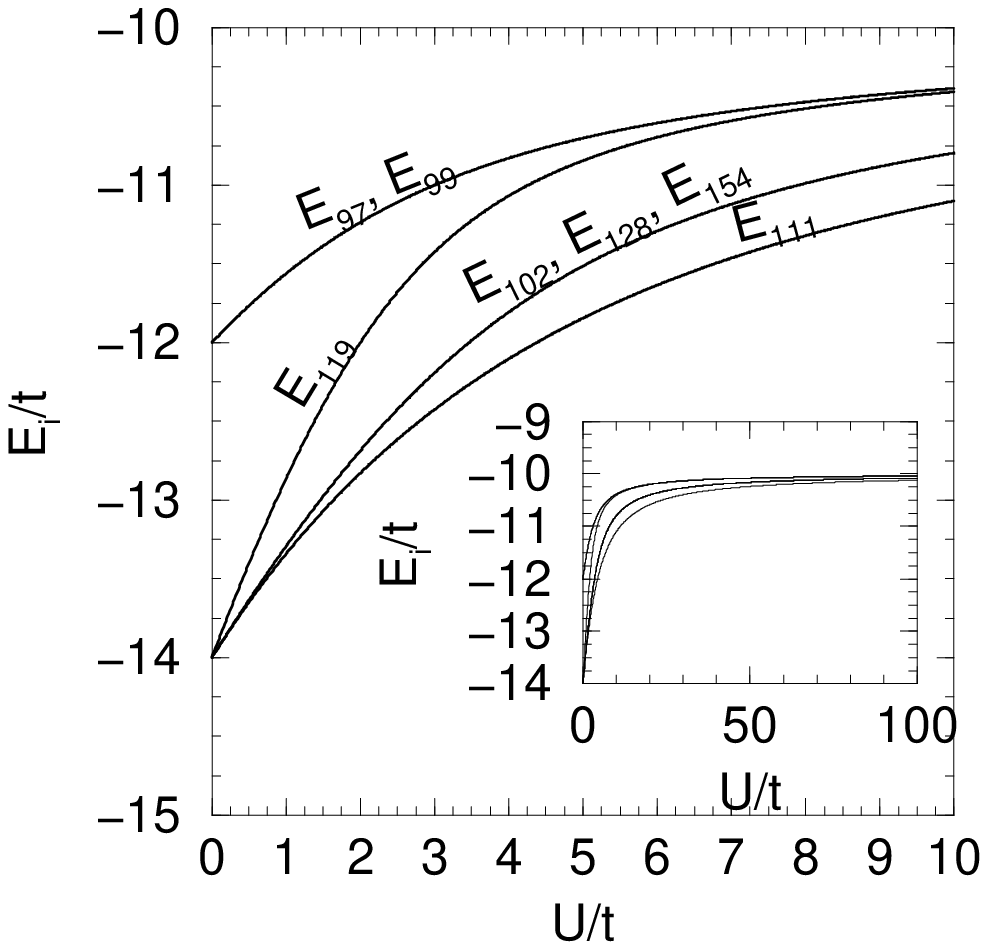,width=8cm}
\end{minipage}\\ \vspace*{\fill} Figure 12
\newpage
\thispagestyle{empty}
\begin{minipage}[t]{12cm}
\psfig{file=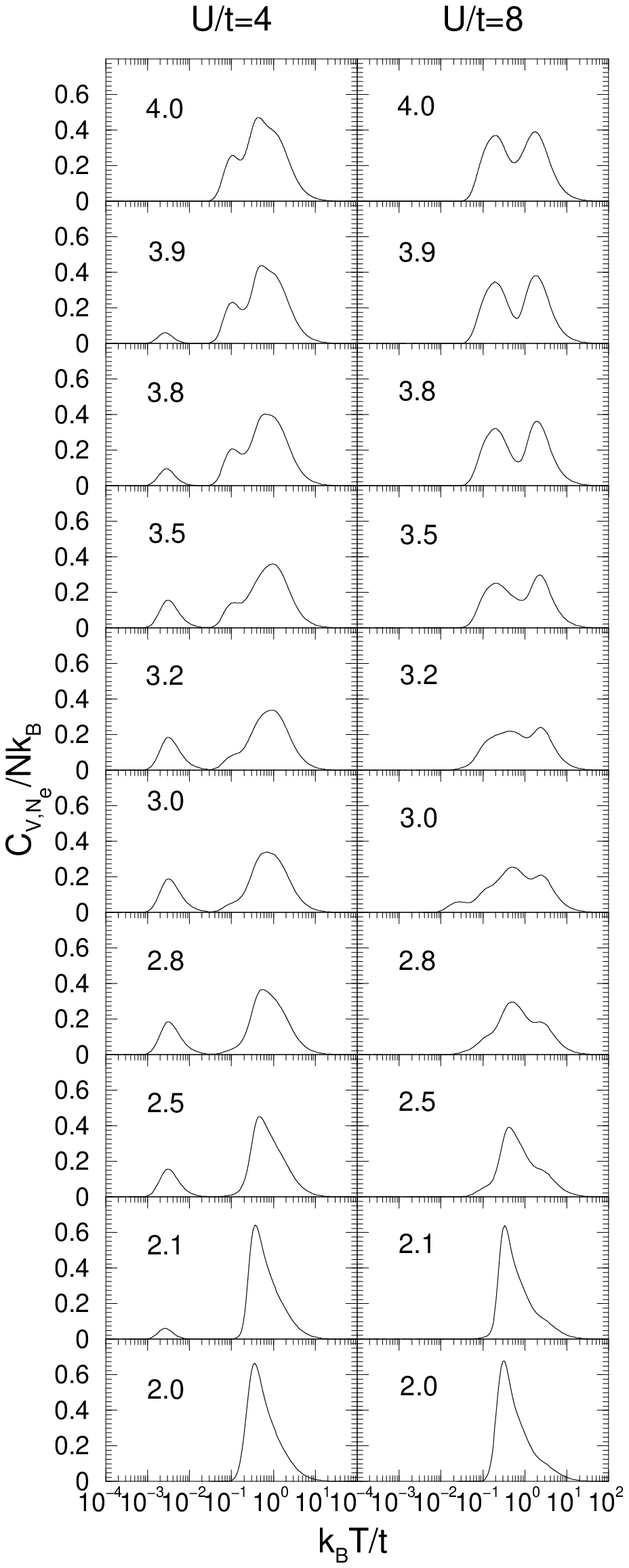,width=7.5cm}
\end{minipage}\\ \vspace*{\fill} Figure 13
\newpage
\thispagestyle{empty}
\begin{minipage}[t]{12cm}
\psfig{file=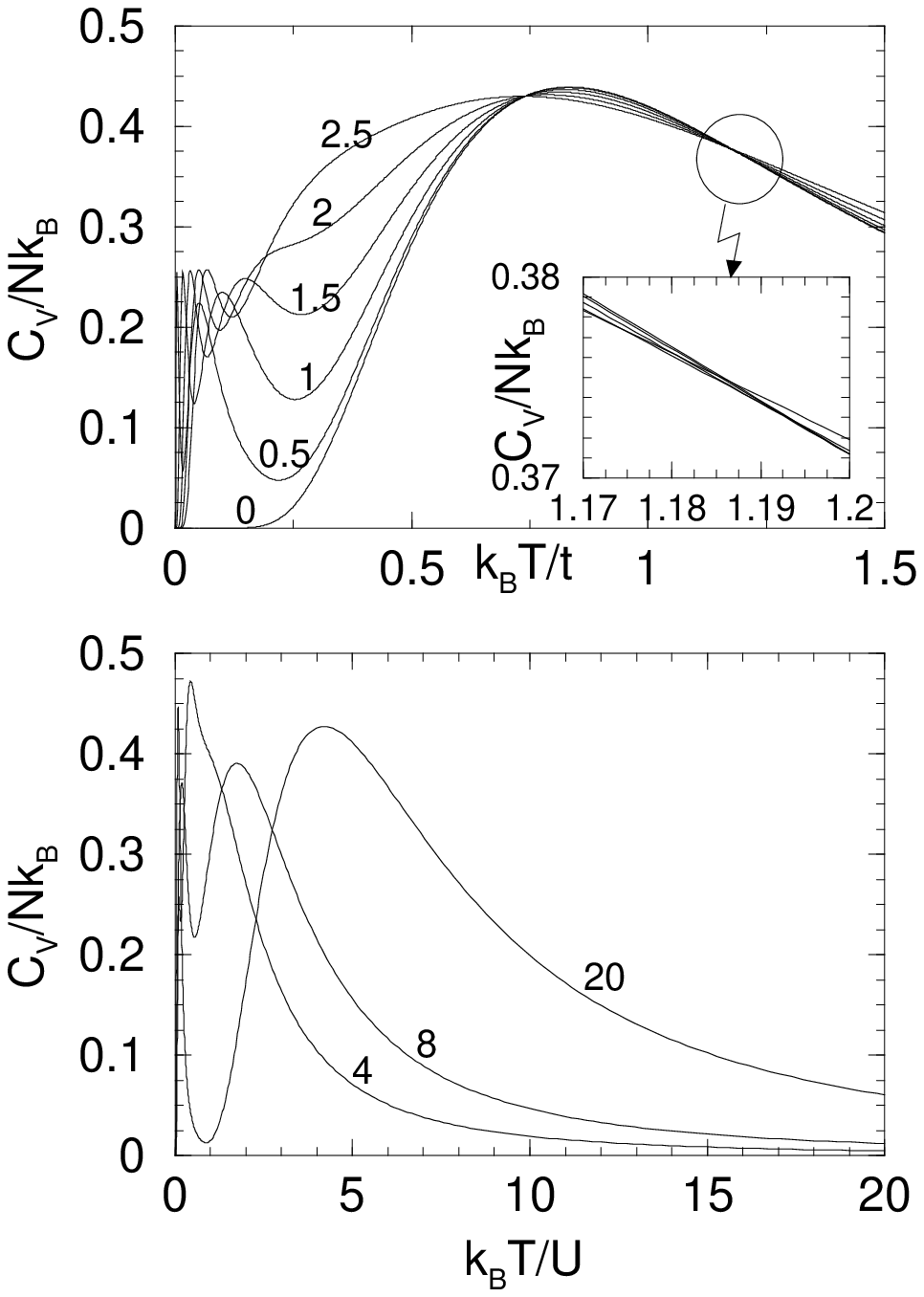,width=8cm}
\end{minipage}\\ \vspace*{\fill} Figure 14
\newpage
\thispagestyle{empty}
\begin{minipage}[t]{12cm}
\psfig{file=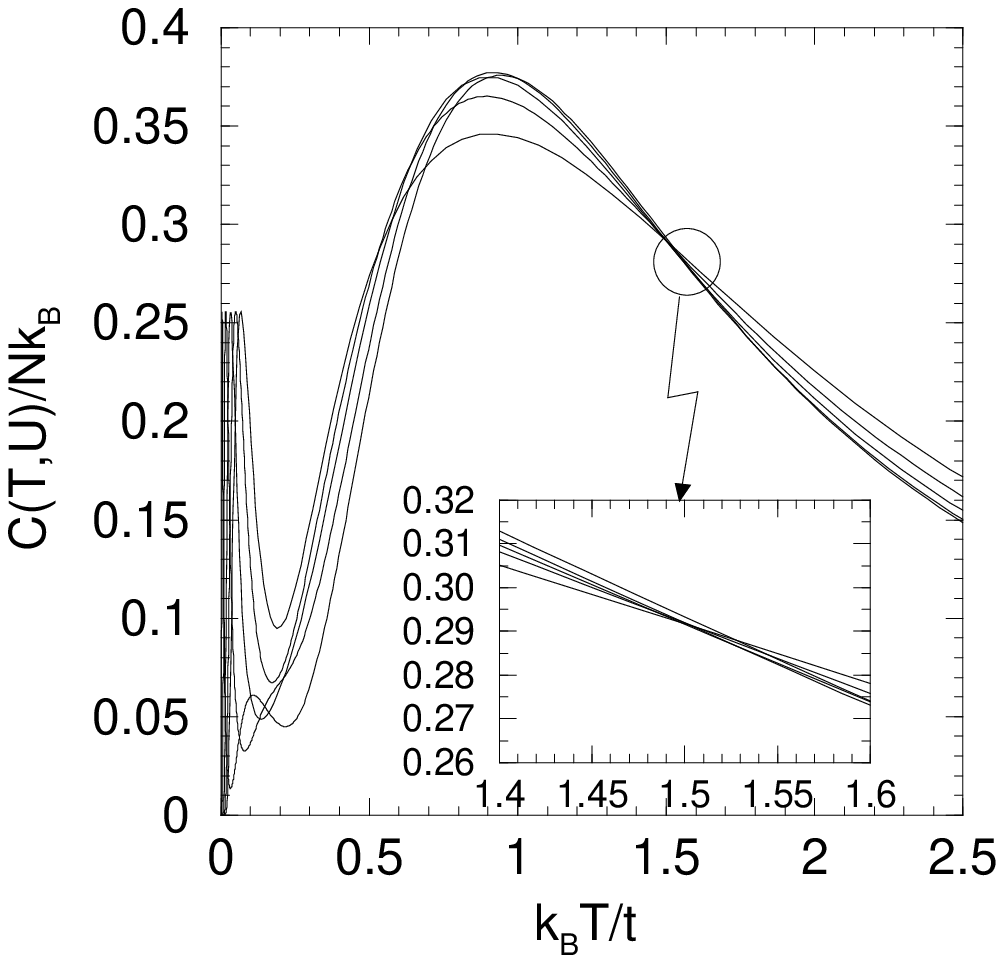,width=8cm}
\end{minipage}\\ \vspace*{\fill} Figure 15
\newpage
\thispagestyle{empty}
\begin{minipage}[t]{12cm}
\psfig{file=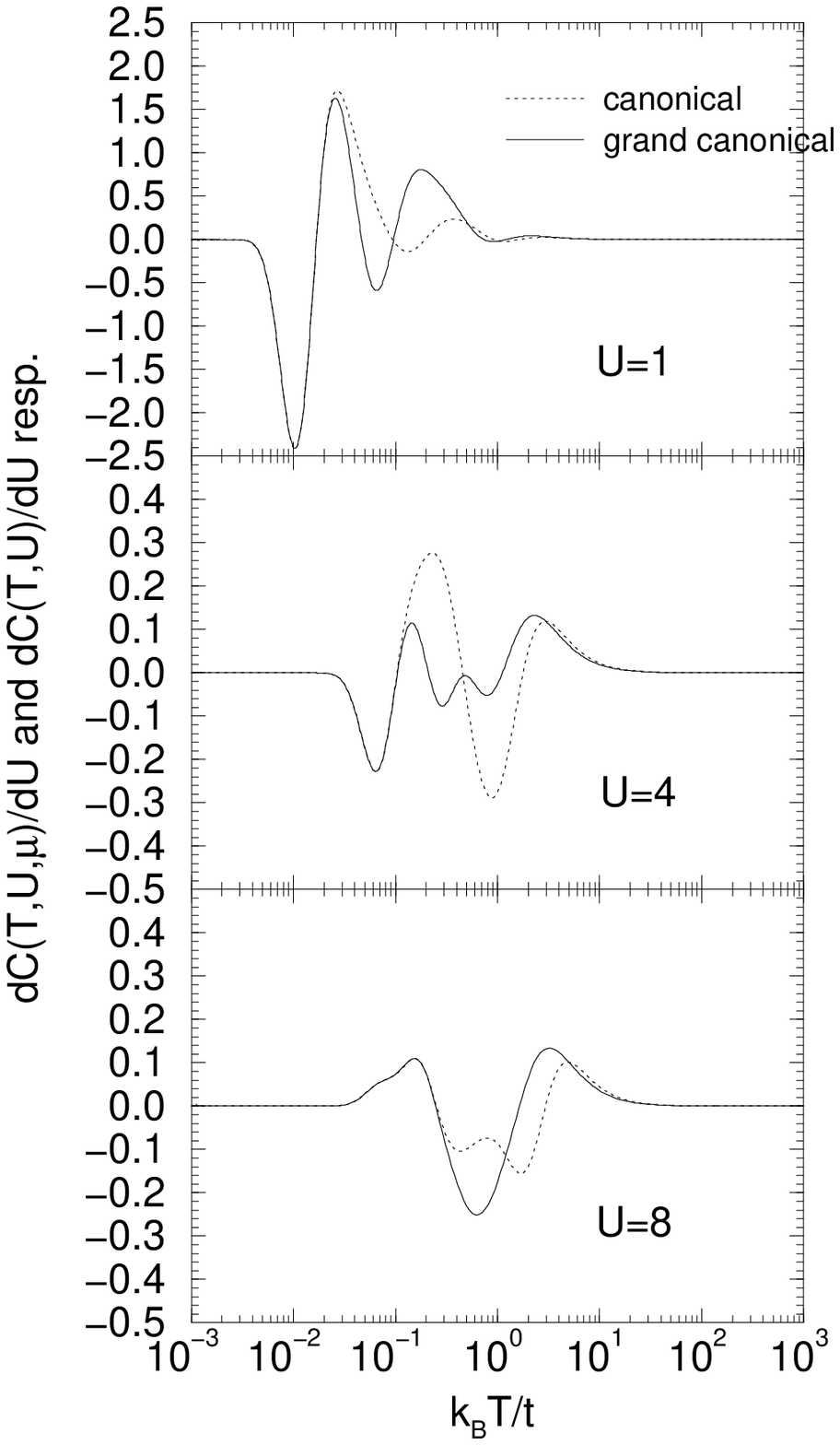,width=8cm}
\end{minipage}\\ \vspace*{\fill} Figure 16
\newpage
\thispagestyle{empty}
\begin{minipage}[t]{12cm}
\psfig{file=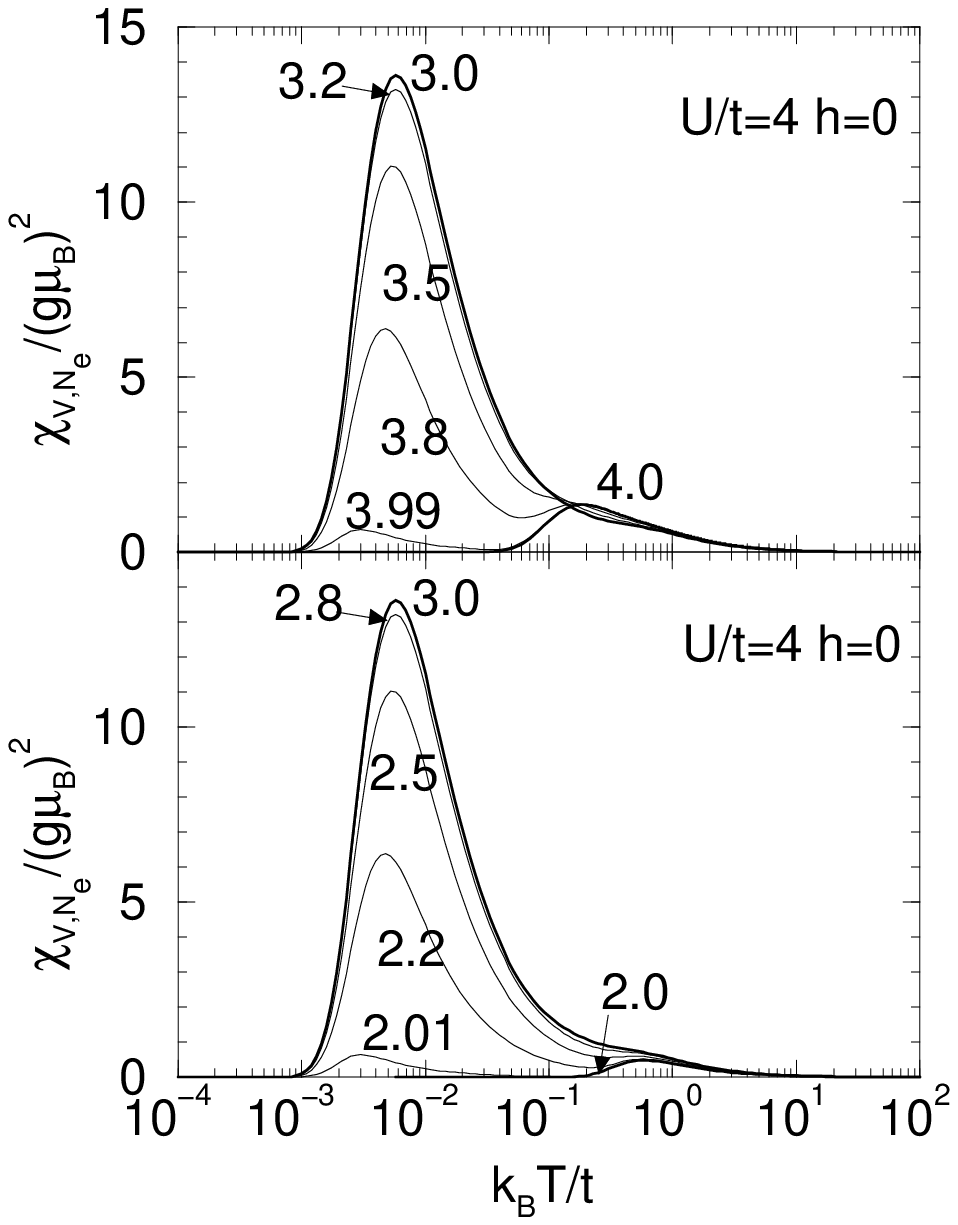,width=8cm}
\end{minipage}\\ \vspace*{\fill} Figure 17
\newpage
\thispagestyle{empty}
\begin{minipage}[t]{12cm}
\psfig{file=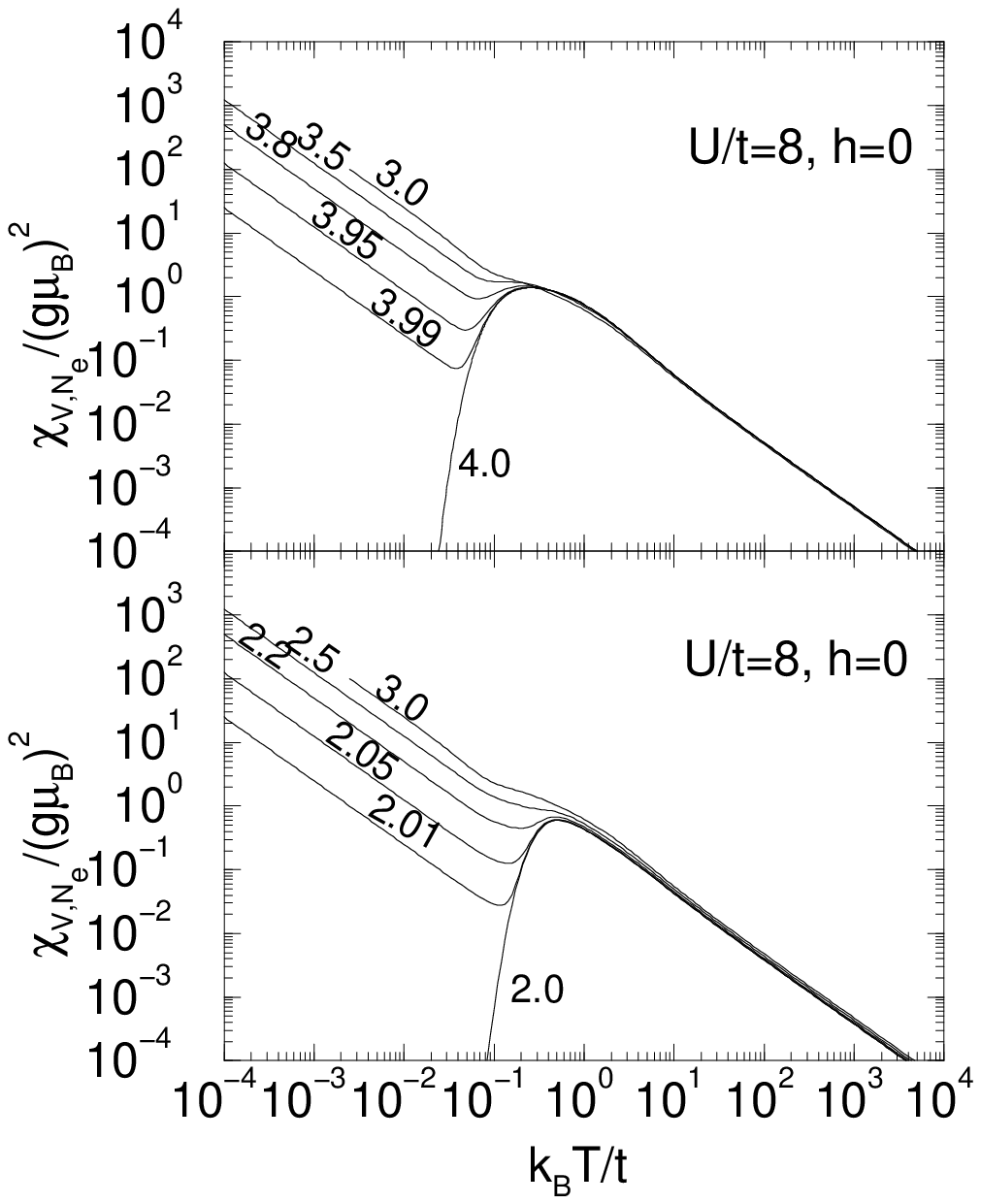,width=8cm}
\end{minipage}\\ \vspace*{\fill} Figure 18
\newpage
\thispagestyle{empty}
\begin{minipage}[t]{12cm}
\psfig{file=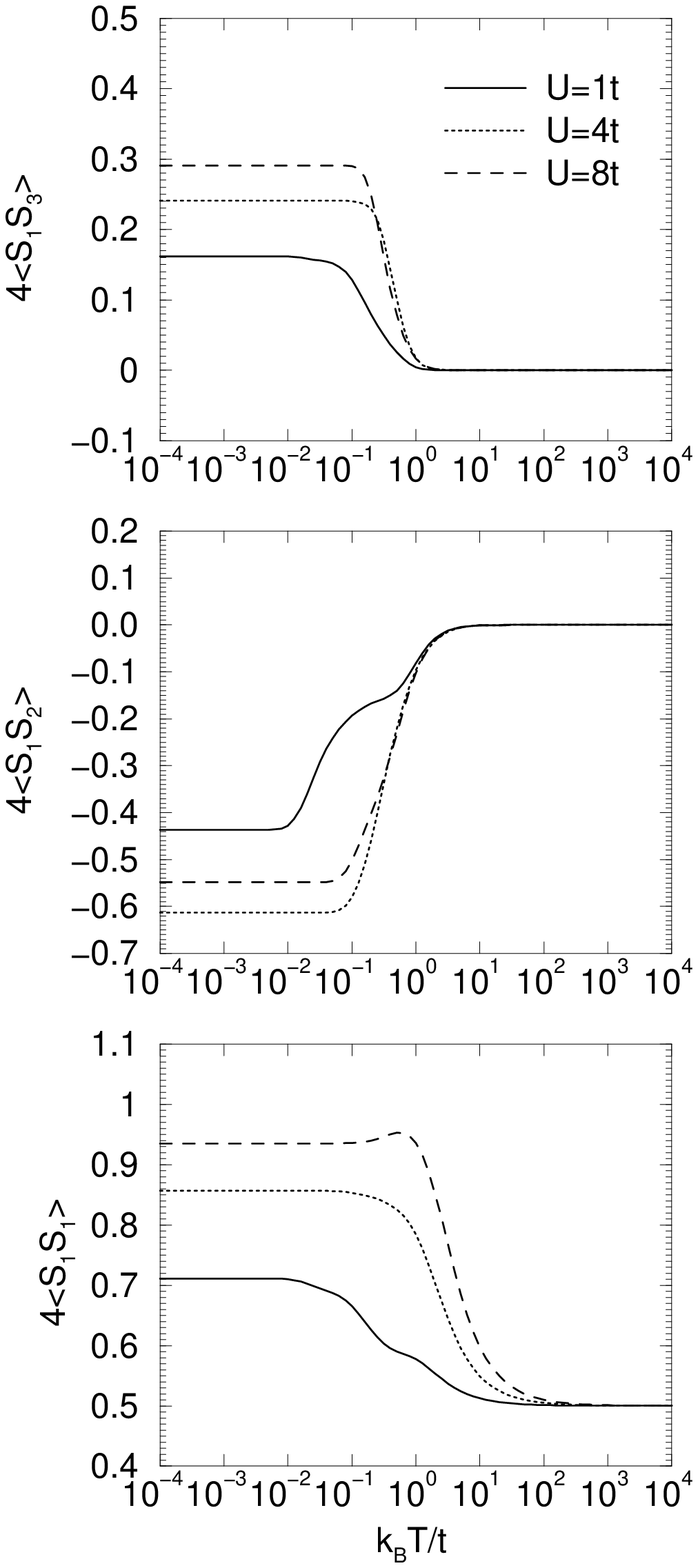,width=8cm}
\end{minipage}\\ \vspace*{\fill} Figure 19
\newpage
\thispagestyle{empty}
\begin{minipage}[t]{12cm}
\psfig{file=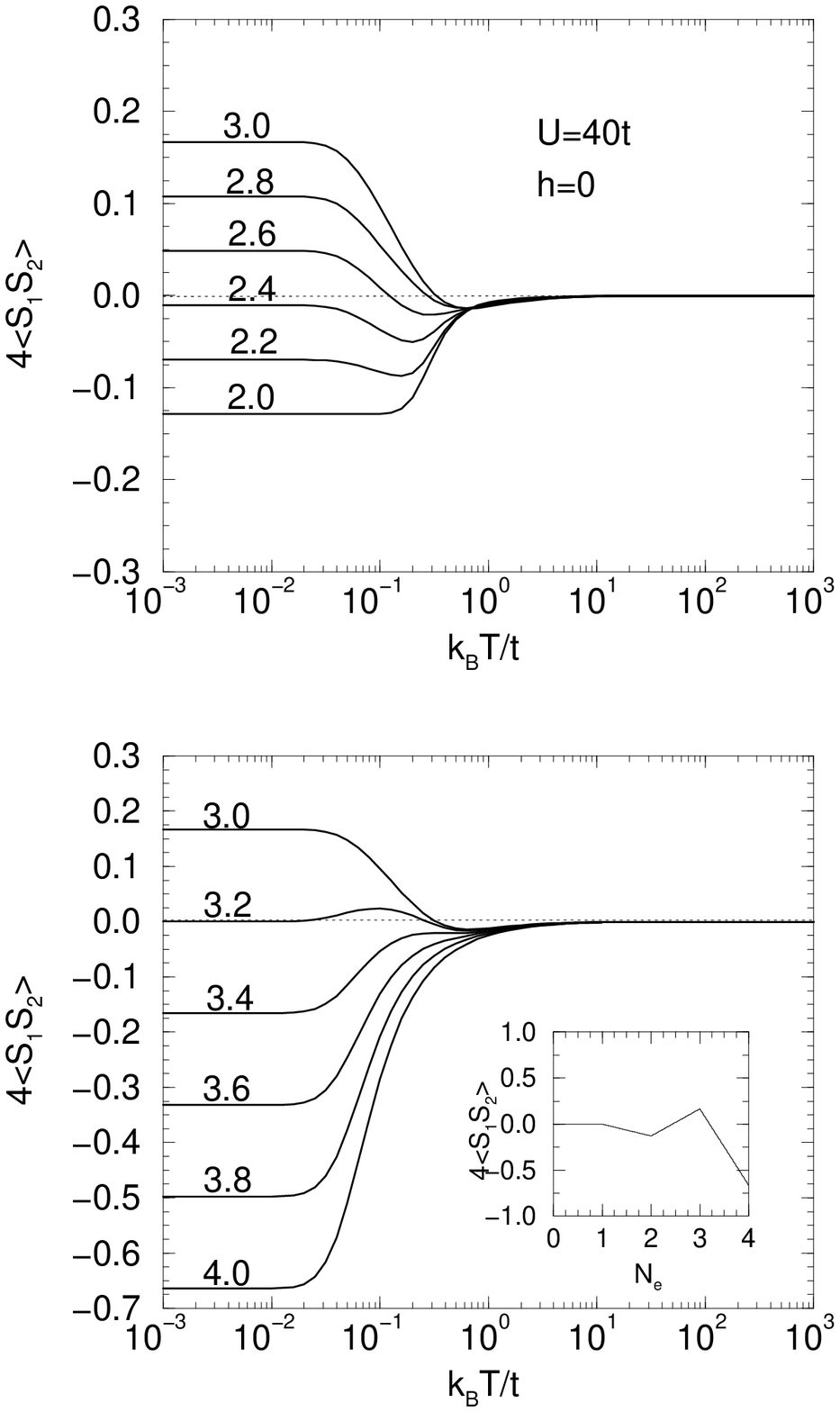,width=8cm}
\end{minipage}\\ \vspace*{\fill} Figure 20
\newpage
\thispagestyle{empty}
\begin{minipage}[t]{12cm}
\psfig{file=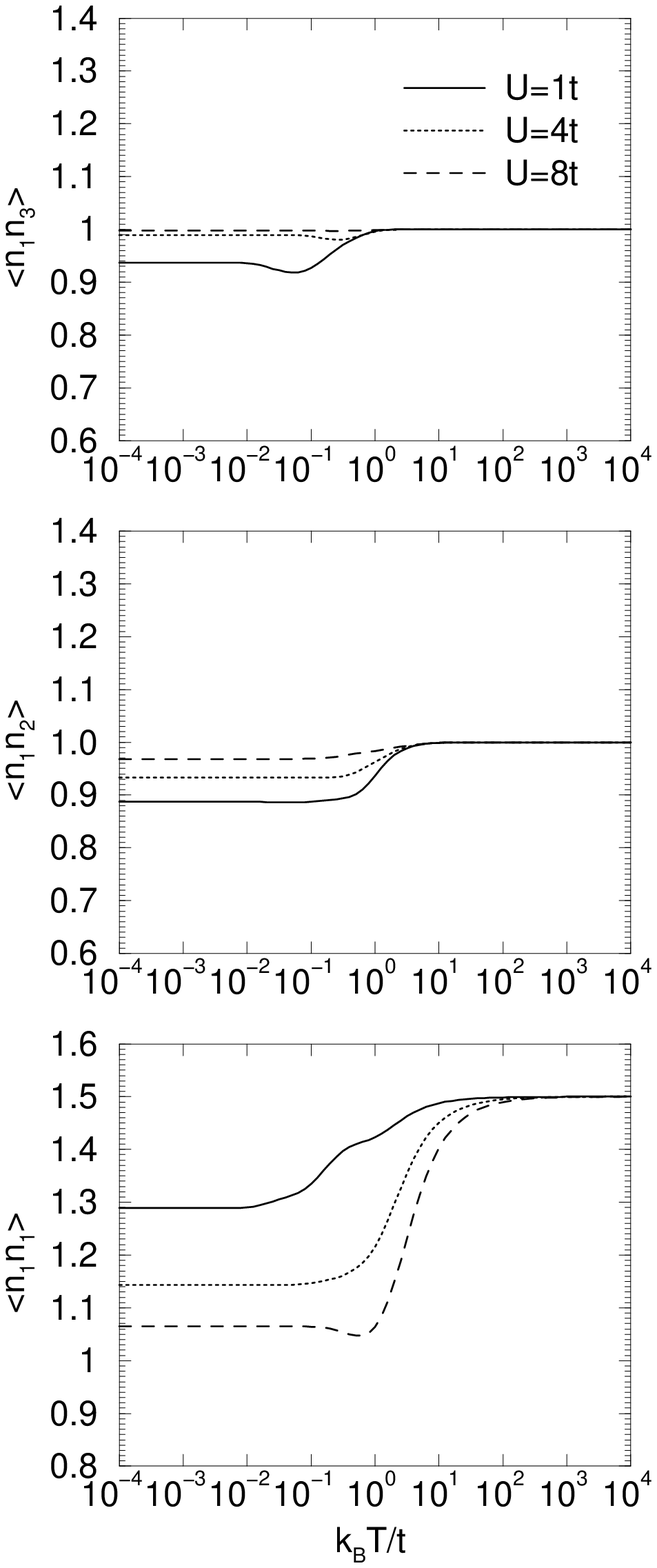,width=8cm}
\end{minipage}\\ \vspace*{\fill} Figure 21
\newpage
\thispagestyle{empty}
\begin{minipage}[t]{12cm}
\psfig{file=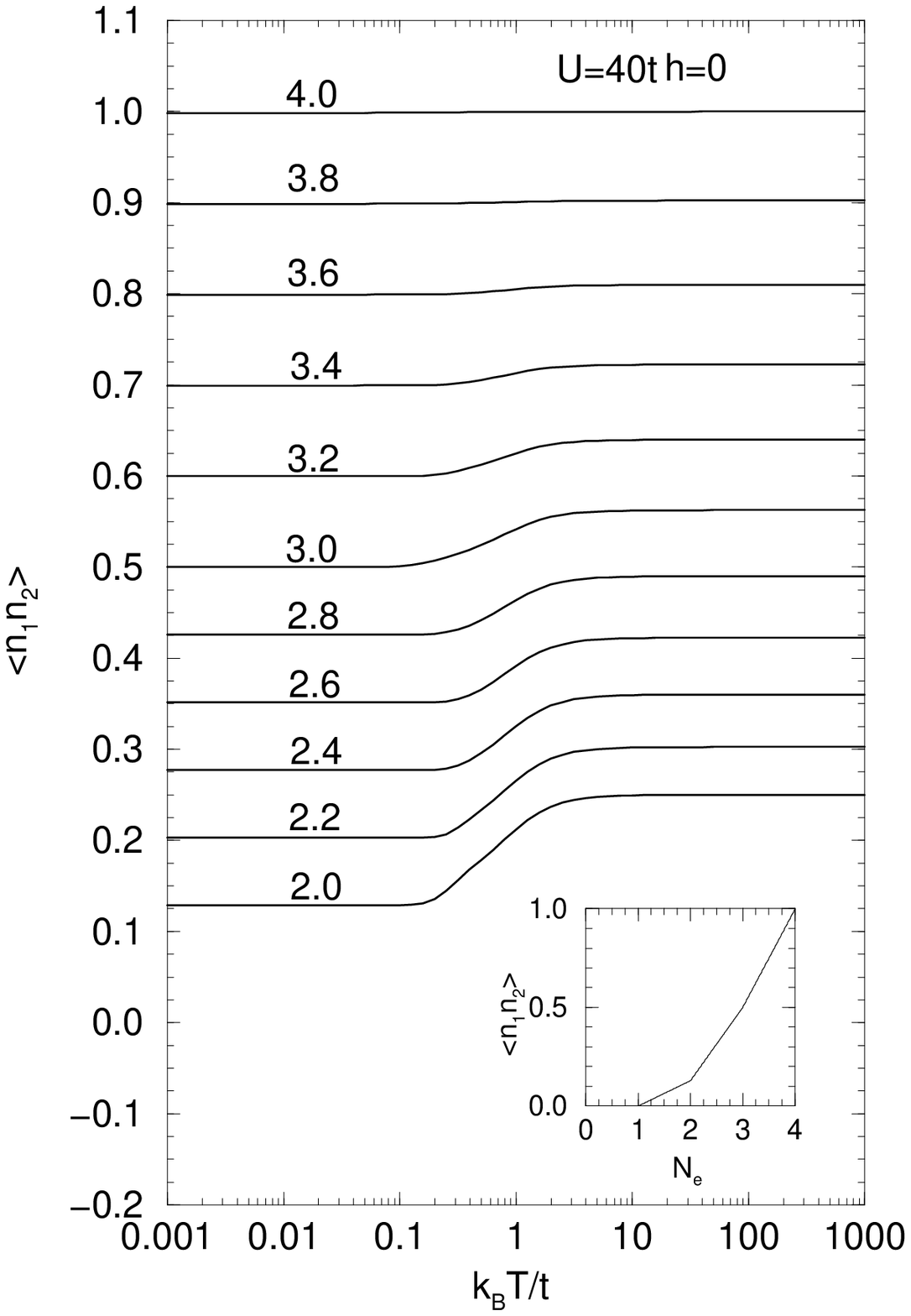,width=8cm}
\end{minipage}\\ \vspace*{\fill} Figure 22
\end{center}
\end{document}